\documentclass[11pt]{article}
\usepackage[defs]{ams}
\usepackage{amsmath}
\usepackage{amssymb,amsthm}
\usepackage{hyperref}
\usepackage{appendix}
\usepackage[all]{xy}
\usepackage[english]{babel}
\usepackage[textwidth=19cm,textheight=24cm]{geometry}
\usepackage{mathrsfs}
\usepackage{mathbbol}

\newtheorem{pro}{Proposition}
\newtheorem{lemma}{Lemma}
\newtheorem{definition}{Definition}
\newtheorem{theorem}{Theorem}
\newtheorem{cor}{Corollary}

\newcommand{\gl}{M_N(\C)}
\newcommand{\E}{W_0}

\newcommand{\I}{\mathbb{I}}
\newcommand{\bS}{\mathbb{S}}

\newcommand{\bt}{\boldsymbol{t}}
\newcommand{\bs}{\boldsymbol{s}}

\newcommand{\Exp}[1]{\operatorname{e}^{#1}}
\newcommand{\diag}{\operatorname{diag}(N,\mathbb C)}

\newcommand{\sg}{\operatorname{sg}}

\newcommand{\cen}{\mathfrak z_\Lambda}
\newcommand{\h}{\mathfrak{h}}

\newcommand{\g}{\mathfrak{g}}

\renewcommand{\r}{\mathcal R}
\renewcommand{\u}{\mathcal U}
\newcommand{\Cc}{\mathcal{C}}

\newcommand{\m}{\mathcal{M}}

\newcommand{\B}{\mathscr B}

\DeclareMathAlphabet{\mathpzc}{OT1}{pzc}{m}{it}

\newcommand{\tb}{\Large\mathpzc b}
\begin{document}

\title{The multicomponent 2D Toda hierarchy:\\ Discrete flows and string equations }

\author{Manuel Ma\~{n}as, Luis Mart\'{\i}nez Alonso, and Carlos \'{A}lvarez Fern\'{a}ndez\\
Departamento de F\'{\i}sica Te\'{o}rica II, Universidad Complutense\\ 28040-Madrid, Spain\\
email: manuel.manas@fis.ucm.es}
\maketitle

\abstract{
The multicomponent 2D Toda hierarchy is analyzed through a factorization problem
associated to an infinite-dimensional group. A  new set of discrete flows is considered and the corresponding Lax and Zakharov--Shabat equations are characterized.  Reductions of block Toeplitz and Hankel bi-infinite matrix types are proposed and studied. Orlov--Schulman operators, string equations and additional symmetries (discrete and continuous) are considered. The continuous-discrete Lax equations are shown to be equivalent to a factorization problem as well as to a set of string equations. A congruence method to derive site independent equations  is presented and used to derive equations in the discrete multicomponent KP sector (and also for its modification)  of the theory as well as  dispersive Whitham equations.}


\section {Introduction}

This paper revisits the multicomponent 2D Toda hierarchy
\cite{ueno-takasaki} from the point of view of the factorization problem
associated to an infinite-dimensional group. Our main motivation  is the
recent discovery [3] of underlying integrable structures of
multicomponent type in the theory  of multiple orthogonal polynomials
which is in turn connected to models of non-intersecting Brownian
motions. Having in mind the  fruitful applications of the Toda hierarchy
to the theory of orthogonal polynomials and to the Hermitian random
matrix model (see for instance \cite{rus}-\cite{mel}), it is expected that
the formalism of multicomponent integrable hierarchies can be similarly
applied to the study and characterization of multiple orthogonal
polynomials and non-intersecting Brownian motions. In particular, the
semiclassical (dispersionless) limit of multicomponent integrable
hierarchies  should be relevant for the analysis of \emph{large $N$})
type  limits, see for instance \cite{mel2}. An important piece of the
technique required for  these applications was recently provided by
Takasaki and Takebe \cite{misgam,takasaki-takebe-ultimo}. Indeed,
they proved that the universal Whitham hierarchy (genus 0 case)
\cite{krichever} can be obtained as a particular dispersionless limit of
the multicomponent KP hierarchy.

The applications of the Toda hierarchy to the characterization of semiclassical limits make an
 essential use of the notion of string equations \cite{rus}-\cite{mel}-\cite{eyn}.
  In recent years the formalism of string equations for dispersionless  integrable hierarchies
  \cite{takasaki-takebe} has been much developed \cite{dispersionless,dispersionless mio} but, to our knowledge,
  a similar formalism for \emph{dispersive} multicomponent integrable hierarchies is not yet available.
  One of the main goals of this paper is to extend the formalism of string equations to multicomponent 2D
  Toda hierarchies. In this sense  the consideration of factorization problems for these hierarchies turns to be of
  great help in order to introduce basic ingredients such as  discrete flows,
  Orlov--Schulman operators and additional symmetries.

The theory of the multicomponent KP hierarchy  is discussed in length
in the papers \cite{kac,tenkroode}, see also \cite{manas2} for its
applications to geometric nets of conjugate type. In
\cite{ueno-takasaki}  it was noticed that $\tau$ functions of a
$2N$-multicomponent KP provide solutions of the $N$-multicomponent
Toda hierarchy.   The introduction of integer parameters in the
multicomponent KP hierarchy goes back to \cite{date} and the
corresponding discrete flows, which are used in two different ways in
\cite{kac,tenkroode}, are essential for the derivation of the
dispersionless Whitham hierarchy from the multicomponent KP
hierarchy \cite{misgam,takasaki-takebe-ultimo},\cite{fin}. In the
present paper we introduce a set of discrete flows for the
multicomponent 2D Toda hierarchy. Its role in the formulation of the
corresponding dispersionless limits will be discussed in length in a
forthcoming paper.


  The layout of the paper is as follows: In \S 1 we introduce a factorization problem in a Lie
  group as the one presented
 in \cite{ueno-takasaki}. This factorization problem is rooted in the ideas  used for the KP case in
\cite{semenov,adler-van moerbeke}, for the so called discrete KP hierarchy. Then  we derive the continuous and discrete Lax equations for the
multicomponent 2D Toda hierarchy. We notice that in our discussion the set of discrete flows, which to our knowledge where not considered before for this hierarchy, are formulated in
 equal footing to the continuous flows.
We also show some examples of members of the hierarchy and,
in particular,  multicomponent equations of Toda type involving  partial difference operators only, or combined partial difference and partial derivatives.
We end this section with the formulation
of  several  classes of reductions of the multicomponent 2D Toda hierarchy involving biinfinite block
Toeplitz and Hankel matrices.  The consideration of these types of reductions is motivated by their relevance in integrable hierarchies such as the infinite  Toda
hierarchy \cite{adler} or the Ablowitz--Ladik lattice hierarchy \cite{caffaso}.
For some reductions we characterize solutions of the hierarchy which are periodic in the discrete variables. Moreover, for the Hankel case we get  generalizations of the bigraded reduction, see \cite{carlet}, associated with extended flows of the
1-component 1D Toda hierarchy \cite{dubrovin}.

In \S 2 we formulate the theory of string equations  for the multicomponent 2D Toda hierarchy. We start by defining  the Orlov--Schulman operator \cite{orlov} and then we derive its Lax equations from the factorization problem introduced in \S 1. We also show how
the Lax equations imply in turn the factorization problem. In this way we stablish the equivalence between the factorization problem and the extended Lax formulation, involving discrete flows and the Orlov--Schulman operator, of the multicomponent 2D Toda hierarchy. Moreover, we also prove the equivalence between the extended  Lax formulation  and a particular type of string equations for the multicomponent 2D Toda hierarchy. This generalizes the result  for the one-component case stablished in \cite{takasaki-takebe}.
Finally, we use the factorization problem and the canonical pair of Lax and  Orlov--Schulman operators to provide a natural formulation of the additional symmetries of the multicomponent 2D Toda hierarchy. As a consequence we characterize the string equations as invariance conditions under additional symmetries.

The paper ends with two appendices. In the first appendix  the congruence
method for deriving $n$-independent  equations  is shown. It is applied to get
the main equations of the discrete multicomponent KP hierarchy: $N$-wave equations, Darboux equations
and  multiquadrilateral lattice equations \cite{quadrilateral}. We also use this method to formulate the dispersive Whitham equations in terms of scalar Lax and Orlov-Schulman opeartors, which constitute the starting point for the discussion of the dispersionless limits of the multicomponent 2D Toda hierarchy. Finally, the second appendix contains the proofs of  the main Propositions of the paper.

\subsection{Lie algebra setting}

In this paper we only consider  formal series expansions in the Lie
group theoretic setup without any assumption on their convergency. We
also remark that along the paper we use the following notations. For
given Lie algebras $\g_1\subset \g_2$, and $X,Y\in\g_2$ then
 $ X=Y+\g_1 $ means $X-Y\in\g_1$. For any Lie groups $G_1\subset G_2$ and $a,b\in G_2$ then
$  a=G_2\cdot b $ stands for $a\cdot b^{-1}\in G_2$. Let
$\{E_{kl}\}_{k,l=1}^N$ be the standard basis
$(E_{kl})_{k'l'}=\delta_{ll'}\,\delta_{kk'}$ of $\gl$ and $\I_N$ denote
the identity matrix in $M_N(\C)$. We also denote the algebra unit as 1.

If $\gl$ denotes the associative algebra  of complex $N\times N$
 matrices we will consider the linear space of sequences
\begin{align*}
\xymatrix@C=2pc@R=1pc{
f: \Z\ar[r]&\gl\\
\quad n\ar@{|->}[r]&f(n).
}
\end{align*}
 The shift operator $\Lambda$ acts on these sequences as
$(\Lambda f)(n):=f(n+1)$. A sequence $X:\Z\to \gl$ acts by left
multiplication in this space of sequences, and therefore we may
consider expressions of the type $X\Lambda^j$, where $X=X(n)$
 is a sequence which acts by left multiplication:
$(  X\Lambda^j)(f)(n):=X(n)\cdot f(n+j)$.

Moreover, defining the product
$(X(n)\Lambda^i)\cdot(Y(n)\Lambda^j):=X(n)Y(n+i)\Lambda^{i+j}$
and extending it linearly we have that the set $\g$ of Laurent
series in $\Lambda$ is an associative algebra, which under the standard commutator is a
Lie algebra.
Observe that $\g$ can be thought either as $M_\Z(M_N(\C))$, i.e. bi-infinite matrices with $M_N(\C)$ coefficients,
 or as $M_N(M_\Z(\C))$, i.e. $N\times N$ matrices with coefficients  bi-infinite matrices.

This Lie algebra has the following important splitting
\begin{gather}\label{splitting}
\g=\g_+\dotplus\g_-,
\end{gather}
where
\begin{align*}
  \g_+&=\Big\{\sum_{j\geq 0}X_j(n)\Lambda^j,\quad X_j(n)\in\gl\Big\},&
  \g_-&=\Big\{\sum_{j< 0}X_j(n)\Lambda^j,\quad X_j(n)\in\gl\Big\},
\end{align*}
are Lie subalgebras of $\g$ with trivial intersection.

\subsection{The Lie group and the factorization problem}
The group of linear invertible elements in $\g$ will be denoted by
$G$ and has $\g$ as its Lie algebra, then the splitting
\eqref{splitting} leads us to consider the following factorization of
$g\in G$
\begin{gather}\label{fac1}
g=g_-^{-1}\cdot g_+, \quad g_\pm\in G_\pm
\end{gather}
where $G_\pm$ have $\g_\pm$ as their Lie algebras. Explicitly, $G_+$
is the set of invertible linear operators  of the
form $\sum_{j\geq 0}g_j(n)\Lambda^j$; while $G_-$ is the set of
invertible linear operators of the form
$1+\sum_{j<0}g_j(n)\Lambda^j$.

An alternative factorization is the Gauss factorization
\begin{gather}\label{facp}
g=\hat g_-^{-1}\cdot \hat g_+, \quad \hat g_\pm\in \hat G_\pm
\end{gather}
where $\hat G_+$ is the set of invertible linear operators of the form
$\hat g_{0,+}(n)+\sum_{j >0}\hat g_j(n)\Lambda^j$; with $\hat
g_{0,+}:\Z\rightarrow \operatorname{GL}(N,\C)$  an invertible upper
triangular matrix,  while $\hat G_-$ is the set of invertible linear
operators of the form $\hat g_{0,-}(n)^{-1}+\sum_{j< 0}\hat
g_j(n)\Lambda^j$ with $\hat g_{0,-}:\Z\rightarrow
\operatorname{GL}(N,\C)$, such that $\hat g_{0,-}=\I_N+A$, being
$A$ a strictly lower triangular matrix in $M_N(\C)$. If the factorization
\eqref{facp} exists then it will also exist \eqref{fac1} by defining
$g_+=\hat g_{0,-}\cdot \hat g_+,\quad g_-=\hat g_{0,-}\cdot\hat g_-$.
The elements $g$ with a factorization \eqref{facp} are said to belong to
the big cell
 \cite{tenkroode}, hence the factorization can be considered only locally. Thus, we will consider elements $g$
 in the big cell so that the factorization \eqref{fac1} holds, avoiding the generation of additional problems
 connected with these local aspects.


 Now we
introduce two sets of indexes, $\mathbb S=\{1,\dots,N\}$ and
$\bar{\mathbb S}=\{\bar 1,\dots,\bar N\}$, of the same cardinality
$N$. In what follows we will use letters $k,l$  and $\bar k,\bar l$ to
denote elements in  $\mathbb S$ and $\bar{\mathbb S}$,
 respectively. Furthermore, we will use letters $a,b,c$ to denote elements in
 $\mathcal S:=\mathbb S\cup\bar{\mathbb
S}$.

We define the following operators $ W_0,\bar  W_0\in G$
\begin{align}
 \label{def:E}  W_0&:=\sum_{k=1}^NE_{kk}\Lambda^{s_k}\Exp{\sum_{j=0}^\infty
 t_{jk}\Lambda^{j}}, \\
\label{def:barE}   \bar W_0&:=\sum_{k=1}^NE_{kk}\Lambda^{- s_{\bar k}}\Exp{\sum_{j=1}^\infty
   t_{j\bar k}\Lambda^{-j}}
\end{align}
where $s_a\in \Z,\, t_{ja}\in \C$ are deformation parameters, that in the sequel
will play the role of discrete and continuous times, respectively.

\paragraph{The factorization problem} Given an element $g\in G$, in the big cell, and a set of
deformation parameters $\bs=(s_a)_{a\in\mathcal
S},\bt=(t_{ja})_{a\in\mathcal S,j\mathbb\in \N_{\sg a}}$, $\N_{+1}=\{0,1,2,\cdots\}$ $\N_{-1}=\{1,2,\cdots\}$,  we  consider the
factorization problem
\begin{gather}
  \label{factorization}
 W_0\cdot g\cdot\bar W_0^{-1}=  S(\bs,\bt)^{-1}\cdot\bar S(\bs,\bt),\quad S\in G_-\text{ and } \bar S\in G_+,
\end{gather}
We will confine our analysis to the \emph{zero charge sector}
\begin{align*}
 |\bs|&:=\sum_{a\in \mathcal S}s_a=0,
\end{align*}
and consider small enough values of the continuous times. Observe that normally, but not always, the $t_{0k}$ times are disregarded.
The reason is the triviality of factorization associated with this deformations. In fact,
if we have a solution of the factorization problem for $t_{0k}=0$, with factors $S_0$ and $\bar S_0$, then the factors corresponding to the factorization
with arbitrary $t_{0k}$ are $S=\exp(\sum_{k=1}^Nt_{0k}E_{kk})S_0\exp(-\sum_{k=1}^Nt_{0k}E_{kk})$ and
$\bar S=\exp(\sum_{k=1}^Nt_{0k}E_{kk})\bar S_0$. The reason to consider them here is due to the reductions we will study later.

At this point we discuss some relevant subalgebras and subgroups which will play an important role hereafter. Firstly, we notice that   an
operator $A=\sum_{j\in\Z} A_j(n)\Lambda^j$ commutes with $\Lambda$
if and only if the coefficients $A_j$ do not depend on $n$. Thus,
the centralizer of $\Lambda$ is
$\cen:=\{A\in\g: [A,\Lambda]=0\}=\Big\{\sum_{j\in\Z}
A_j\Lambda^j, A_j\in\gl\Big\}$.
Observe that $\cen\subset \g$ is a Lie subalgebra 
as now $\Lambda$ commutes
with the matrix coefficients of the Laurent expansions. Another interpretation is that we have block bi-infinite
Toeplitz  or Laurent operators \cite{toeplizt}.

A particular Abelian subalgebra $\h$ of $\cen$ is given by the centralizer of  $\C\{\Lambda,E_{kk}\}_{k=1}^N$; i.e,
$\h:=\{A\in\g: [A,\Lambda]=[A,E_{kk}]=0,k=1,\dots,N\}=\Bigl\{\sum_{j\in\Z} A_j\Lambda^j, A_j\in\diag\Bigr\}$
where $\diag$ is the subalgebra of diagonal matrices of $\gl$.
 Thus, $\h$ is the set of Laurent series in $\Lambda$ with diagonal $n$-independent coefficients.
There are two important subgroups: $G_-\cap\cen=\{1+c_1\Lambda^{-1}+c_2\Lambda^{-2}+\cdots,
 c_j\in\gl\}$ and $G_+\cap\cen=\{\bar c_0+\bar c_1\Lambda+\bar
c_2\Lambda^2+\cdots,\quad \bar C_0\in \text{GL}(N,\C),\; \bar
c_j\in\gl, j\geq 1\}$.
Finally, we have the corresponding Abelian Lie subgroups
$H:=G\cap \h$ and $H_\pm:=G_\pm\cap\h$
and $ W_0,\bar  W_0$ takes values in $H$.

We shall denote by $n\in\g$ the multiplication operator by the sequence $\{n \I_N\}_{n \in\Z}$; i.e.
\begin{gather}\label{mus}
n \{X(n)\}_{n\in\Z}=\{nX(n)\}_{n\in \Z}.
\end{gather}
Observe that $[\Lambda,n]=\Lambda$
and that for any $X\in\g$ we have
$X=\sum_{\substack{j\in \Z\\ i \geq 0}}X_{ij} n^i\Lambda^j$, $ X_{ij}\in
\gl$.
This expansion follows from the assumption that
    $X_j(n)=X_{0j}+X_{ij}n+\cdots$.
The set of operators commuting with $\Lambda,n$ and $E_{kk}$, $k=1,\dots,N$ is given by
$
\{A\in\g: [A,\Lambda]=[A,n]=[A,E_{kk}]=0,k=1,\dots,N\}=
\diag$.

\section{ Lax  and Zakharov--Shabat equations}

\subsection{Dressing procedure.  Lax and $C$ operators}
We now introduce  important elements for the sequel of this paper
\begin{definition}\label{baker}
 We   define the dressing operators $W,\bar W$ as follows
\begin{align}
\label{def:baker}W&:=S\cdot W_0,& \bar W&:=\bar S\cdot \bar  W_0,
\end{align}
\end{definition}

In terms of these dressing operators the factorization problem\eqref{factorization} in $G$ reads
\begin{gather}
  \label{facW}
  W\cdot g=\bar W
\end{gather}

Observe that the expansions of the factors $S,\bar S$
\begin{gather}
\label{expansion-S}
\begin{aligned}
S&=\I_N+\varphi_1(n)\Lambda^{-1}+\varphi_2(n)\Lambda^{-2}+\cdots\in G_-,\\
\bar
S&=\bar\varphi_0(n)+\bar\varphi_1(n)\Lambda+\bar\varphi_2(n)\Lambda^{2}+\cdots\in
G_+.
\end{aligned}
\end{gather}
Sometimes we will use the notation
\begin{align*}
  \beta&:=\varphi_1,& \Exp{\phi}&:=\bar\varphi_0.
\end{align*}
 We have the following expressions
\begin{gather}\label{baker expansion}
\begin{aligned}
 W&=(\I_N+\varphi_1(n)\Lambda^{-1}+\varphi_2(n)\Lambda^{-2}+\cdots)\cdot\Big(\sum_{k=1}^NE_{kk}
 \Lambda^{s_k}
\exp\Big(\sum_{j=0}^\infty t_{jk}\Lambda^j\Big)\Big)\\
\bar W&=(\bar\varphi_0(n)+\bar\varphi_1(n)\Lambda+\bar\varphi_2(n)\Lambda^{2}+\cdots)\cdot
\Big(\sum_{k=1}^NE_{kk}
 \Lambda^{- s_{\bar k}}
\exp\Big(\sum_{j=1}^\infty  t_{j\bar k}\Lambda^{-j}\Big)\Big)
\end{aligned}
\end{gather}

Other important objects are
\begin{definition}\label{lax-C}
 The Lax  operators $L,\bar L,C_{kl},\bar C_{kl},\Cc_{kl},\bar \Cc_{kl}\in\g$ are given by
\begin{align}
\label{Lax}  L&:=W\cdot\Lambda\cdot W^{-1}, & \bar L&:=\bar W\cdot\Lambda\cdot \bar W^{-1}, \\
\label{C} C_{kl}&:=W\cdot E_{kl}\cdot W^{-1},& \bar C_{kl}&:=\bar
W\cdot E_{kl}\cdot \bar W^{-1}\\
 \label{Cc} \Cc_{kl}&:=S\cdot
E_{kl}\cdot S^{-1},& \bar \Cc_{kl}&:=\bar S\cdot E_{kl}\cdot \bar
S^{-1}.
\end{align}
\end{definition}
Notice that in the above definitions of $L,\bar L,C_{kk}$ and $\bar
C_{kk}$ ---as $ W_0$, $\bar W_0\in H$--- we may replace the dressing
operators $W$ and $\bar W$ by  $S$ and $\bar S$, respectively. A
straightforward calculations yields
\begin{pro}\label{lax-C2}
  \begin{enumerate}
  \item The following relations holds
  \begin{align*}
  C_{kl}&=L^{s_k-s_l}\exp(\sum_{j=0}^\infty
(t_{jk}-t_{jl})L^j))\Cc_{kl},\\
\bar C_{kl}&=\bar L^{- s_{\bar k}+ s_{\bar l}}\exp(\sum_{j=1}^\infty
( t_{j\bar k}- t_{j\bar l})\bar L^{-j}))\bar\Cc_{kl}.
\end{align*}
\item
The Lax operators have the following expansions
\begin{gather}\label{lax expansion}
\begin{aligned}
 L&=\Lambda+u_1(n)+u_2(n)\Lambda^{-1}+\cdots,&
\bar L^{-1}&=\bar u_0(n)\Lambda^{-1}+\bar u_1(n)+\bar u_2(n)\Lambda+\cdots, \\
\Cc_{kl}&=E_{kl}+C_{kl,1}(n)\Lambda^{-1}+C_{kl,2}(n)\Lambda^{-2}+\cdots,&
\bar \Cc_{kl}&=\bar C_{kl,0}(n)+\bar C_{kl,1}(n)\Lambda+\bar
C_{kl,2}(n)\Lambda^{2}+\cdots.
\end{aligned}
\end{gather}
\item These   operators fulfill
\begin{gather}\label{identity}
\begin{aligned}
 \mathbb{I}_N&=\sum_{k=1}^NC_{kk},&\mathbb{I}_N&=\sum_{k=1}^N\bar C_{kk},&
\end{aligned}\\
\label{LCrel}\begin{aligned}
C_{kl}C_{k'l'}&=\delta_{lk'}C_{kl'},& C_{kl}L&=LC_{kl},\\
 \bar C_{kl}\bar
C_{k'l'}&=\delta_{lk'}\bar C_{kl'},& \bar C_{kl}\bar L&=\bar L\bar
C_{kl},
\end{aligned}
\end{gather}
  \end{enumerate}
\end{pro}

\subsection{Lax and Zakharov--Shabat equations}

In this section we will use the factorization problem \eqref{facW} to derive  two sets of equations:  Lax equations and Zakharov--Shabat equations, and we will show they all are equivalent.
Let us first introduce some convenient notation
\begin{definition}\label{def:omega}
\begin{enumerate}
\item\begin{align*}
&  \partial_{ja}:=\frac{\partial}{\partial t_{ja}},
\end{align*}
\item The \emph{zero-charge} shifts
$T_{K}$ for $K=(a,b)$ are defined as follows
\begin{align*}
s_a&\to s_a+1,& s_b&\to s_b-1,
\end{align*}
and all the others discrete variables remain unchanged.
\item
\begin{align*}
\theta_{ja}&:=
\partial_{ja}W_0\cdot W_0^{-1},&\bar\theta_{ja}&:=\bar
\partial_{ja}\bar W_0\cdot \bar W_0^{-1},\\
q_{K}&:=T_K W_0\cdot W_0^{-1},& \bar q_{K}&:=T_K\bar W_0\cdot \bar
W_0^{-1},\quad K=(a,b).
\end{align*}
\item
\begin{align}\label{RU}
\begin{aligned}
C_{aa}&:=W\pi_aW^{-1} &\bar C_{aa}&:=\bar W\bar \pi_a\bar W^{-1},\\
\r_{ja}&:= W\theta_{ja}W^{-1},& \bar \r_{ja}&:=  \bar W\bar \theta_{aj}\bar W^{-1},&
\u_K&:=Wq_KW^{-1},&
\bar\u_K&:=\bar W\bar q_K\bar W^{-1}.
\end{aligned}\end{align}
\item
\begin{align}
\label{omega}
\begin{aligned}
B_{ja}&:=\r_{ja}-(\r_{ja}-\bar \r_{ja})_-=\bar \r_{ja}+(\r_{ja}-\bar \r_{ja})_+\in\g,& (\r_{ja}-\bar\r_{ja})_\pm\in\g_\pm,\\
\omega_K&:=(\u_k\cdot\bar \u_K^{-1})_-\cdot\u_K=(\u_k\cdot\bar \u_K^{-1})_+\cdot\bar\u_K\in G, & (\u_K\bar\u_K^{-1})_\pm\in G_\pm.
\end{aligned}
\end{align}
\end{enumerate}
\end{definition}
Notice that if
\begin{align*}
\pi_a&:=\begin{cases}  E_{kk}, &
a=k\in\mathbb S,\\
0,& a\in\bar{\mathbb S},
\end{cases} &
\bar \pi_a&:=\begin{cases} 0, &
a\in\mathbb S,\\
 E_{kk},& a\in\bar{\mathbb S} \text{ and $a=\bar k$ for some $k\in\mathbb S$}.
\end{cases}\end{align*}
we can write
\begin{align*}
\theta_{ja}&= \pi_a\Lambda^j,&\bar\theta_{ja}&=\bar
\pi_a\Lambda^{-j},\\
q_{K}&=\I_N+\pi_a(\Lambda-\I_N)+\pi_b(\Lambda^{-1}-\I_N),& \bar
q_{K}&=\I_N+\bar\pi_a(\Lambda^{-1}-\I_N)+\bar\pi_b(\Lambda-\I_N),\quad
K=(a,b).
\end{align*}

 Observe that all the shift operators preserve the zero
charge sector and form a commutative group
\begin{align}
&T_KT_{K'}=T_{K'}T_K,\label{T-abelian}\\
&T_{(a,b)}T_{(b,a)}=\operatorname{id},\label{T-inverse}
\end{align}
satisfying the following cohomological relations
\begin{align}
  &T_{(a,b)}T_{(b,c)}T_{(c,a)}=\text{id}.\label{T-cohomological}
\end{align}

\begin{pro}\label{T-pro}
The relations \eqref{T-abelian}-\eqref{T-cohomological} are equivalent to
\begin{align}
\label{T-relations}
T_{(a,b)}T_{(b,c)}=T_{(b,c)}T_{(a,b)}=T_{(a,c)}
\end{align}
where $T_{(a,a)}=\operatorname{id}$.
\end{pro}
\begin{proof}
See Appendix B.
\end{proof}
  Also notice that $  B_{jk}=(C_{kk}L^j)_+$,  $B_{j\bar k}=(\bar C_{kk}\bar L^{-j})_-$
and that \eqref{RU} and \eqref{omega} gives
\begin{align}\label{omega-a}
  \omega_K=\pi_a\Lambda+a_K+\bar a_K\Lambda^{-1},
\end{align}
for some matrix sequences  $a_K(n)$ and $\bar a_K(n)$.

The factorization problem \eqref{factorization} implies that
the  partial differential equations
 \begin{align}
 \label{tjk}
 \partial_{ja}W\cdot W^{-1}=\partial_{ja}S\cdot S^{-1}+S\cdot \theta_{ja}\cdot S^{-1}
&=\partial_{ja}\bar S\cdot \bar S^{-1}+\bar S\cdot
\bar\theta_{ja}\cdot \bar S^{-1}=\partial_{ja}\bar W\cdot \bar
W^{-1},
\end{align}
and partial difference equations
\begin{gather}
T_K W\cdot W^{-1}=
T_K S\cdot q_K\cdot S^{-1}=
T_K\bar S\cdot\bar q_K\cdot \bar S^{-1}=T_K\bar W\cdot\bar W^{-1}
\label{Tkl}
\end{gather}
hold.

 From the previous proposition we derive the following linear systems for the dressing operators and Lax equations for the Lax operators, and its compatibility conditions
\begin{theorem}\label{pro:integrable hierarchies}
\begin{enumerate}
\item
The dressing operators are subject to
\begin{align}
\label{eq:bakerjk} \partial_{ja}W&=B_{ja}\cdot W,& \partial_{ja}\bar W&=B_{ja}\cdot\bar W,\\
 \label{eq:dbakerjk}    T_KW&=\omega_K\cdot W,&    T_K\bar W&=\omega_K\cdot\bar W.
   \end{align}
 \item  The  Lax equations
   \begin{align}
\label{laxtjk}
  \partial_{ja} L&= [B_{ja},L],&\partial_{ja} \bar L&= [B_{ja},\bar L],&
 \partial_{ja} C_{kk}&= [B_{ja},C_{kk}],&\partial_{ja} \bar C_{kk}&= [B_{ja},\bar C_{kk}],
\\
\label{dlaxtkl}T_K L&=\omega_K\cdot L\cdot
\omega_K^{-1},& T_K\bar L&=\omega_K\cdot \bar L\cdot
\omega_K^{-1},&
T_K C_{kk}&=\omega_K\cdot C_{kk}\cdot \omega_K^{-1},& T_K\bar
C_{kk}&=\omega_K\cdot \bar C_{kk}\cdot \omega_K^{-1},
\end{align}are satisfied.
\item The following Zakharov--Shabat equations hold
\begin{gather}
\label{zs} \partial_{ja} B_{ib}- \partial_{ib}B_{ja}+[B_{ib},B_{ja}]=0,\\
\label{Omega-omega} T_KB_{ja}=\partial_{ja} \omega_K\cdot
\omega_K^{-1}+\omega_K\cdot B_{ja}\cdot\omega_K^{-1},\\
T_K\omega_{K'}\cdot\omega_K=T_{K'}\omega_{K}\cdot\omega_{K'}.
\label{con-omega1}
\end{gather}
\end{enumerate}
\end{theorem}

\begin{proof}
\begin{enumerate}
\item
First, observe that \eqref{tjk} implies
 $ \partial_{ja} S\cdot  S^{-1}+\r_{ja}=  \partial_{ja}\bar S\cdot \bar S^{-1}+\bar\r_{ja},$
and therefore
$  \partial_{ja}S\cdot S^{-1}=-(\r_{ja}-\bar\r_{ja})_-\in\g_-$ and
 $ \partial_{ja}\bar S\cdot \bar S^{-1}=(\r_{ja}-\bar\r_{ja})_+\in\g_+$
so that using again \eqref{tjk} we get
\begin{gather}\label{eq:B1}
  \partial_{ja}W\cdot W^{-1}=-(\r_{ja}-\bar\r_{ja})_-+\r_{ja}=B_{ja}=
  (\r_{ja}-\bar\r_{ja})_++\bar\r_{ja}=\partial_{ja}\bar W\cdot \bar  W^{-1}
\end{gather}

 Equation \eqref{Tkl} implies
\begin{align}\label{us}
T_KW\cdot W^{-1}=T_K S\cdot   q_K \cdot S^{-1}=T_K S\cdot S^{-1}\cdot   \u_K=T_K\bar S\cdot \bar S^{-1}\cdot \bar \u_K =T_K\bar S\cdot \bar q_K\cdot \bar S^{-1}=T_K\bar W\cdot \bar W^{-1}
\end{align}
so that
$(T_K S\cdot S^{-1})^{-1}\cdot (T_K\bar S\cdot \bar S^{-1})=\u_K\cdot\bar \u_K^{-1}$
and we conclude $T_K S\cdot S^{-1}=(\u_K\cdot\bar \u_K^{-1})_-\in G_-$ and
$T_K\bar S\cdot \bar S^{-1}=(\u_K\cdot\bar \u_K^{-1})_+\in G_+$
which introduced back in \eqref{us} gives
\begin{multline}\label{omega W}
 T_KW\cdot W^{-1}= T_K S\cdot   q_K \cdot S^{-1}=(\u_K\cdot\bar \u_K^{-1})_-\cdot   \u_K=\omega_K\\=(\u_K\cdot\bar \u_K^{-1})_+\cdot \bar \u_K =T_K\bar S\cdot \bar q_K\cdot \bar S^{-1}=T_K\bar W\cdot \bar W^{-1}.
\end{multline}

\item From the definition \eqref{Lax} we get
\begin{align*}
  \partial_{ja}L&=[\partial_{ja} W\cdot W^{-1},L],&   \partial_{ja}\bar L&=[\partial_{ja}\bar W
  \cdot \bar W^{-1},\bar L],\\
  \partial_{ja} C_{kk}&=[\partial_{ja}\bar W\cdot\bar W^{-1},C_{kk}],&
  \partial_{ja}\bar C_{kk}&=
  [\partial_{ja}\bar W\cdot\bar W^{-1},\bar C_{kk}],\\
  T_K L&=(T_KW\cdot W^{-1})\cdot L\cdot(T_KW\cdot W^{-1})^{-1}, &   T_K\bar L&=(T_K\bar W\cdot \bar W^{-1})\cdot \bar L\cdot(T_K\bar W\cdot \bar W^{-1})^{-1},\\
    T_K C_{kk}&=(T_KW\cdot W^{-1})\cdot  C_{kk}\cdot(T_KW\cdot W^{-1})^{-1}, &   T_K\bar  C_{kk}&=(T_K\bar W\cdot \bar W^{-1})\cdot \bar  C_{kk}\cdot(T_K\bar W\cdot \bar W^{-1})^{-1}.
\end{align*}
and using \eqref{eq:bakerjk} and \eqref{eq:dbakerjk} we find \eqref{laxtjk} and \eqref{dlaxtkl}, respectively.

\item The compatibility of \eqref{eq:bakerjk} and \eqref{eq:dbakerjk}
imply \eqref{zs}-\eqref{con-omega1}
\end{enumerate}
\end{proof}

Observe also that the trivial flows $t_{0k}$, $k=1,\dots,N$
 are immediately integrated and if $L_0$, $\bar L_0$, $C_{kl}$ and
$\bar C_{kl}$ are the Lax and $C$ operators corresponding to $t_{0k}=0$ for arbitrary $t_{k0}$ we only need to conjugate these operators with
$\exp(\sum_{k=1}^NE_{kk}t_{k0})$.

The compatibility conditions \eqref{zs}-\eqref{con-omega1} for
operators $B_{ja}$ and $\omega_K$ formally imply the local existence
of a matrix potential $\xi$ such that
$B_{ja}=\partial_{ja}\xi\cdot\xi^{-1}$ and
$\omega_K=T_K\xi\cdot\xi^{-1}$; here, the potential $\xi$ is a map to
the Lie group $G$ depending on the variables $\{t_{ja}, s_a\}$ i.
Moreover, any operator $\xi$ generates a gauge transformation so that
 $ B_{ja}\rightarrow\partial_{ja}\xi\cdot\xi^{-1}+\xi\cdot B_{ja}\cdot\xi^{-1}$ and
$  \omega_K\rightarrow T_K\xi\cdot\omega_K\cdot\xi^{-1}$,
providing  new solutions of \eqref{zs}-\eqref{con-omega1}.

\begin{pro}\label{T-pro2}
 The relations  \begin{align}\label{con-omega}
  \big(T_{(a,b)}\omega_{(b,c)}\big)\omega_{(a,b)}=\big(T_{(b,c)}\omega_{(a,b)}\big)\omega_{(b,c)}=
\omega_{(a,c)}.
\end{align}
 and  the compatibility conditions \eqref{con-omega1} are equivalent.
\end{pro}
\begin{proof}
See Appendix B.
\end{proof}

We have seen that the Lax equations \eqref{laxtjk}-\eqref{dlaxtkl}
and Zakharov--Shabat equations \eqref{zs}-\eqref{con-omega1}
appear as consequence of the factorization problem. The compatibility conditions for the Lax
 equations are satisfied if the Zakharov--Shabat equations hold. It is a standard fact in the theory of Integrable Systems that by construction the Lax equations imply the Zakharov--Shabat equations and
 therefore the system is compatible. In \cite{ueno-takasaki} is proven this fact for the differential equations
 (not the difference nor difference-differential equations) involved in the multicomponent 2D Toda hierarchy,
  that is that \eqref{laxtjk} $\Rightarrow$ \eqref{zs}. Here we give an extended proof in order to include the continuous-discrete and discrete-discrete cases.

\begin{pro}\label{lax then zs}
  Let $\{L,C_{kk}\}_{k=1}^N\subset\g$ and
  $ \{\bar L,\bar C_{kk}\}_{k=1}^N\subset\g$
   be two sets, composed each of them of  commuting operators, consider
    functions $\r_{ja}\in\g$, $\u_K\in G$  of $L,C_{11},\dots,C_{kk}$
     and $\bar \r_{ja}\in\g$, $\bar \u_K\in G$  of
     $\bar L,\bar C_{11},\dots,\bar C_{kk}$,
      and define $B_{ja}$ and $\omega_K$ according to
       \eqref{omega}, then the Lax equations \eqref{laxtjk}
        and \eqref{dlaxtkl} imply the Zakharov--Shabat equations \eqref{zs}-\eqref{con-omega1}.
\end{pro}
\begin{proof}
See Appendix B.
\end{proof}

\subsection{The multicomponent Toda equations}
Here we write down some of the nonlinear partial differential-difference equations appearing as a consequence of the factorization problem \eqref{facW}. From \eqref{eq:B1} and \eqref{omega W},
taking into account that   $S\in G_-$ and $\bar S\in G_+$, we deduce the following
\begin{cor}
  We have the expressions
\begin{gather}\label{exp-omega}
\begin{aligned}
B_{1a}&=\pi_a\Lambda+U_a+\bar U_a\Lambda^{-1},\\
  \omega_{K}&:=\pi_a\Lambda+a_K+\bar a_K\Lambda^{-1},\quad K=(a,b),
  \end{aligned}
\end{gather}
where the coefficients have the alternative expressions
\begin{gather}\label{exp-omega1}
\begin{aligned}
  U_a&:=\beta(n)\pi_a-\pi_a\beta(n+1)=\begin{cases}
\partial_{1a}(\Exp{\phi(n)})\cdot\Exp{-\phi(n)},& a\in\mathbb S\\
0, &a\in\bar{\mathbb S}\\
  \end{cases}\\
 \bar U_a&= \Exp{\phi(n)}\bar\pi_a\Exp{-\phi(n-1)}=\begin{cases}0,& a\in\mathbb S,\\
   \partial_{1a}\beta(n),& a\in\bar{\mathbb S},
 \end{cases}\\
a_K&:=\I_N-\pi_a-\pi_b+T_{K}\beta(n)\pi_a-
\pi_a\beta(n+1)=\begin{cases}
\Exp{T_K\phi(n)}\cdot(\I_N-\bar\pi_b)\cdot\Exp{-\phi(n)}, & a\in\mathbb S,\\\I_N-\pi_b, & a\in\bar{\mathbb S},
\end{cases}\\
\bar a_K&:=\Exp{T_{K}\phi(n)}\bar\pi_a\Exp{-\phi(n-1)}=
\begin{cases}0,& a\in\mathbb S,\\
 T_K\beta(n)(\I_N-\pi_b)-(\I_N-\pi_b)\beta(n)+ \pi_b, & a\in\bar{\mathbb S}.
\end{cases}
\end{aligned}
\end{gather}
\end{cor}
From \eqref{exp-omega1} we deduce the following set of nonlinear
partial differential-difference equations
\begin{align}\left\{
\begin{aligned}
 \beta(n)E_{kk}-E_{kk}\beta(n+1)&=\partial_{1k}(\Exp{\phi(n)})\cdot\Exp{-\phi(n)},\\
\partial_{1\bar k}\beta(n)&=\Exp{\phi(n)}E_{kk}\Exp{-\phi(n-1)},\\
    T_{(k,b)}\beta(n)E_{kk}-E_{kk}\beta(n+1)+\I_N-E_{kk}-\pi_b&=\Exp{T_{(k,b)}\phi(n)}\cdot(\I_N-\bar\pi_b)\cdot\Exp{-\phi(n)},
\\
T_{(\bar k,b)}\beta(n)(\I_N-\pi_b)-(\I_N-\pi_b)\beta(n)+ \pi_b&=\Exp{T_{(\bar k,b)}\phi(n)}\cdot E_{kk}\cdot\Exp{-\phi(n-1)}.
\end{aligned}\right.
\label{eq:multitoda}
\end{align}
These equations constitute what we call the multicomponent Toda equations. Observe that if we cross the two first equations we get
\begin{align*}
  \partial_{1\bar k'}\big(\partial_{1k}(\Exp{\phi(n)})\cdot\Exp{-\phi(n)}\big)=
 \Exp{\phi(n)}E_{k'k'}\Exp{-\phi(n-1)}E_{kk}- E_{kk}\Exp{\phi(n+1)}E_{k'k'}\Exp{-\phi(n)}
\end{align*}
which is the matrix extension of the 2D Toda equation, which appears for $N=1$:
\begin{align*}
  \partial_{1}\partial_{\bar 1}(\phi(n))=
 \Exp{\phi(n)-\phi(n-1)}- \Exp{\phi(n+1)-\phi(n)}.
\end{align*}
If in the last equation we set $b=\bar l\in\bar{\mathcal S}$ we have
\begin{align*}
  \Delta_{(\bar k,\bar l)}\beta(n)&=\Exp{T_{(\bar k,\bar l)}\phi(n)}\cdot E_{kk}\cdot\Exp{-\phi(n-1)}.
\end{align*}
which when considered simultaneously with the first gives
\begin{align*}
  \Delta_{(\bar k',\bar l)}\big(\partial_{1k}(\Exp{\phi(n)})\cdot\Exp{-\phi(n)}\big)=
  \Exp{T_{(\bar k',\bar l)}\phi(n)}\cdot E_{k'k'}\cdot\Exp{-\phi(n-1)}E_{kk}-E_{kk}\Exp{T_{(\bar k',\bar l)}\phi(n+1)}\cdot E_{k'k'}\cdot\Exp{-\phi(n)}
\end{align*}
which is a Toda type equation. A completely discrete equation appears, for example, when crossing the two last equations, i.e.
\begin{multline*}
  \Delta_{(\bar k',\bar l)}\big(\Exp{T_{(k,b)}\phi(n)}\cdot(\I_N-\bar\pi_b)\cdot\Exp{-\phi(n)}\big)=
   T_{(k,b)}\big(\Exp{T_{(\bar k',\bar l)}\phi(n)}\cdot E_{k'k'}\cdot\Exp{-\phi(n-1)}\big)E_{kk}
   -E_{kk} \Exp{T_{(\bar k',\bar l)}\phi(n+1)}\cdot E_{k'k'}\cdot\Exp{-\phi(n)}.
\end{multline*}
So forth and so on we may get a set of continuous-discrete set of Toda
type equations. Finally, observe that when $N=1$ we only have the
shift $T_{(s_1,s_{\bar 1})}$ which corresponds to a shift $n\to n+1$.

\subsection{Block Toeplitz/Hankel reductions}
We now consider some  reductions of the multicomponent 2D Toda
hierarchy. In the first place we discuss
 an extension of the
periodic reduction  \cite{ueno-takasaki} and the bigraded reduction
\cite{carlet} to the multicomponent case, which we call Toeplitz/Hankel
reduction. Finally we discuss an extension of the 1 dimensional
reduction discussed in \cite{ueno-takasaki}. These reductions are
relevant when we work with semi-infinite cases, as in the construction
of families of bi-orthogonal and orthogonal matrix polynomials, to be
published elsewhere.

Given a set $\{\ell_a\}_{a\in \mathcal S}\subset\Z$
 we seek for initial conditions $g$ satisfying
\begin{align}\label{periodic g}
  g\cdot\Big(\sum_{k=1}^NE_{kk}\Lambda^{-\ell_{\bar k}}\Big)=
  \Big(\sum_{k=1}^NE_{kk}\Lambda^{\ell_k}\Big)\cdot g.
\end{align}
The relation \eqref{periodic g} gives the following  constraints over the Lax operators
\begin{align}\label{reduction}
\sum_{k=1}^N C_{kk} L^{\ell_{k}j}&=\sum_{k=1}^N\bar C_{kk} \bar L^{-\ell_{\bar k}j}
\end{align}
for any $j\in\Z$.
To proceed further in the analysis of these reductions we define the sets
\begin{align*}
  \bS_\pm&:=\{a\in\bS: \pm\ell_a>0\}, & \bS_0&:=\{a\in\bS: \ell_a=0\},&\bar\bS_\pm&:=\{a\in\bar\bS: \pm\ell_a>0\}, & \bar\bS_0&:=\{a\in\bar\bS: \ell_a=0\},
\end{align*}
so  that
$  \bS=\bS_+\cup\bS_0\cup\bS_-$, and $\bar \bS=\bar\bS_+\cup\bar\bS_0\cup\bar\bS_-$.
\begin{pro} \label{pro:reduction}
  If \eqref{reduction} holds then we have
  \begin{enumerate}
  \item The dressing operators are subject to
  \begin{align}\label{periodic W}
  \begin{aligned}
       \Big(  \sum_{a\in \bS_+\cup\bS_0\cup\bar\bS_+ }\partial_{j\ell_a,a}\Big)(W)&=W\sum_{k=1}^NE_{kk}\Lambda^{j\ell_k},&
   \Big(  \sum_{a\in \bS_+\cup\bS_0\cup\bar\bS_+ }\partial_{j\ell_a,a}\Big)(\bar W)&=\bar W\sum_{k=1}^NE_{kk}\Lambda^{-j\ell_{\bar k}},\\
     \Big(  \sum_{a\in \bS_-\cup\bS_0\cup \bar\bS_-}\partial_{j|\ell_a|,a}\Big)(W)&=W\sum_{k=1}^NE_{kk}\Lambda^{-j\ell_k},&
   \Big(  \sum_{a\in \bS_-\cup\bS_0\cup \bar\bS_-}\partial_{j|\ell_a|,a}\Big)(\bar W)&=\bar W\sum_{k=1}^NE_{kk}\Lambda^{j\ell_{\bar k}},
  \end{aligned}
  \end{align}
  for $j>0$.
    \item
   The Lax operators are invariant:
    \begin{align}\label{periodic L}
    \begin{aligned}
  &\Big(  \sum_{a\in \bS_+\cup\bS_0\cup\bar\bS_+ }\partial_{j\ell_a,a}\Big)(L)=  \Big(\sum_{a\in \bS_+\cup\bS_0\cup\bar\bS_+ }\partial_{j\ell_a,a}\Big)(\bar L)=0,\\
    &\Big( \sum_{a\in \bS_-\cup\bS_0\cup \bar\bS_-}\partial_{j|\ell_a|,a}\Big)(L)=
      \Big(\sum_{a\in \bS_-\cup\bS_0\cup \bar\bS_-}\partial_{j|\ell_a|,a}
    \Big)(\bar L)=0,
\end{aligned}
\end{align}
where $j>0$.
  \end{enumerate}
  Moreover, if
\begin{align*}
  \sum_{a\in\mathcal S}\ell_a=0;
\end{align*}
then,
\begin{enumerate}
\item The dressing operators fulfill
\begin{align}\label{periodic W 2}
\begin{aligned}
W(s_1+\ell_1,\dots,s_N+\ell_N,s_{\bar 1}+\ell_{\bar 1},\dots,s_{\bar N}+\ell_{\bar N})&=
  W(s_1,\dots,s_N,s_{\bar 1},\dots,s_{\bar N})\sum_{k=1}^NE_{kk}\Lambda^{\ell_k},\\
  \bar W(s_1+\ell_1,\dots,s_N+\ell_N,s_{\bar 1}+\ell_{\bar 1},\dots,s_{\bar N}+\ell_{\bar N})&=
   \bar W(s_1,\dots,s_N,s_{\bar 1},\dots,s_{\bar N})\sum_{k=1}^NE_{kk}\Lambda^{-\ell_{\bar k}}.
\end{aligned}
\end{align}
  \item The Lax operators are periodic
\begin{align}\label{Lperiod}
 \begin{aligned}
  L(s_1+\ell_1,\dots,s_N+\ell_N,s_{\bar 1}+\ell_{\bar 1},\dots,s_{\bar N}+\ell_{\bar N})&=
  L(s_1,\dots,s_N,s_{\bar 1},\dots,s_{\bar N}),\\ \bar L(s_1+\ell_1,\dots,s_N+\ell_N,s_{\bar 1}+\ell_{\bar 1},\dots,s_{\bar N}+\ell_{\bar N})&=
  \bar L(s_1,\dots,s_N,s_{\bar 1},\dots,s_{\bar N}).
\end{aligned}
\end{align}
\end{enumerate}

  \end{pro}
To prove this we need the
\begin{lemma}\label{reduction invariance}
If \eqref{reduction} holds then for $j>0$ we have
\begin{align}\label{invariance}
\begin{aligned}
    \sum_{a\in \bS_+\cup\bS_0\cup\bar \bS_+ }B_{j\ell_a,a}
  &=\sum_{k=1}^N C_{kk} L^{\ell_{k}j}=
  \sum_{k=1}^N\bar C_{kk} \bar L^{-\ell_{\bar k}j},\\
  \sum_{a\in \bS_-\cup\bS_0\cup \bar\bS_-}B_{j|\ell_ a|,a}&=\sum_{k=1}^N C_{kk} L^{-\ell_{k}j}=
  \sum_{k=1}^N\bar C_{kk} \bar L^{\ell_{\bar k}j}.
\end{aligned}
\end{align}
\end{lemma}
\begin{proof}
  The projection on $\g_+$ of $A=\sum_{k=1}^N C_{kk} L^{\ell_{k}j}$, $j>0$, is $\sum_{a\in \bS_+\cup\bS_0}B_{j\ell_a,a}$ while the projection on $\g_-$ of   $A=\sum_{k=1}^N\bar C_{kk} \bar L^{-\ell_{\bar k}j}$, $j>0$, is   $\sum_{a\in\bar \bS_+ }B_{j\ell_a,a}$. The first formula is just $A=A_++A_-$. The second formula follows in a similar way when $j<0$.
\end{proof}
Now we proceed with
\begin{proof}[Proof of Proposition \ref{pro:reduction}]
 Equations \eqref{periodic W} and \eqref{periodic L} follow from the previous lemma and Theorem \ref{pro:integrable hierarchies}. To deduce  \eqref{periodic W 2} and \eqref{Lperiod} we argue as follows.  If
\begin{align*}
  \sum_{a\in\mathcal S}\ell_a=0
\end{align*}
 the periodicity follows from the factorization problem
\begin{align*}
    S\cdot W_0\cdot \Big(\sum_{k=1}^NE_{kk}\Lambda^{\ell_k}\Big)\cdot g=S\cdot W_0\cdot g\cdot \Big(\sum_{k=1}^NE_{kk}\Lambda^{-\ell_{\bar k}}\Big)=
  \bar S\cdot\bar W_0 \cdot\Big(\sum_{k=1}^NE_{kk}\Lambda^{-\ell_{\bar k}}\Big)
\end{align*}
by observing that
\begin{align*}
W_0(s_1+\ell_1,\dots,s_N+\ell_N)&=W_0(s_1,\dots,s_N) \sum_{k=1}^NE_{kk}\Lambda^{\ell_k},\\
\bar W_0(s_{\bar 1}+\ell_{\bar 1},\dots,s_{\bar N}+\ell_{\bar N})&=
\bar W_0(s_{\bar 1},\dots,s_{\bar N}) \sum_{k=1}^NE_{kk}\Lambda^{-\ell_{\bar k}},\\
\end{align*}
and recalling the uniqueness property of the factorization problem we deduce the periodicity condition for the solutions
\begin{align*}
  S(s_1+\ell_1,\dots,s_N+\ell_N,s_{\bar 1}+\ell_{\bar 1},\dots,s_{\bar N}+\ell_{\bar N})&=
  S(s_1,\dots,s_N,s_{\bar 1},\dots,s_{\bar N}),\\ \bar S(s_1+\ell_1,\dots,s_N+\ell_N,s_{\bar 1}+\ell_{\bar 1},\dots,s_{\bar N}+\ell_{\bar N})&=
  \bar S(s_1,\dots,s_N,s_{\bar 1},\dots,s_{\bar N}),
\end{align*}
which imply
\eqref{periodic W 2}
and \eqref{Lperiod}.
\end{proof}

Now we justify the name of this reduction.  If we write $g=\sum_{j\in\Z}g_j(n)\Lambda^j$,  and think of it as an element in $M_N(M_\Z(\C))$, i.e.
$g=\sum_{k_1,k_2=1}^Ng_{k_1k_2}E_{k_1k_2}$ and $g_{k_1k_2}=\sum_{j\in\Z}g_{j,k_1k_2}(n)\Lambda^j$ then \eqref{periodic g} gives
\begin{align}
\label{periodic}
g_{j,k_1k_2}(n)=g_{j+\ell_{k_1}+\ell_{\bar k_2},k_1k_2}(n-\ell_{k_1}).
\end{align}
If $\ell_{k_1}+\ell_{\bar k_2}=0$, then $g_{j,k_1k_2}$ is a $|\ell_{k_1}|$-periodic function in $n$.
If this period is 1,  we get that $g_{k_1k_2}$ is a bi-infinite Toeplitz or Laurent matrix .
We will see that in the general case we are dealing with block Toeplitz \cite{toeplizt} and block Hankel \cite{hankel} bi-infinite matrices.
\begin{definition}
  Given a block matrix $\Omega=(\Omega_{i,j})_{i,j\in\Z}$ made up with
$p\times q$-blocks $\Omega_{i,j}$ we say that $\Omega$ is a block Toeplitz matrix  if
$\Omega_{i+1,j+1}=\Omega_{i,j}$ and a block Hankel matrix  if $\Omega_{i+1,j-1}=\Omega_{i,j}$.
\end{definition}

\begin{pro}\label{matrix structure}
  The condition \eqref{periodic} implies for $g_{k_1k_2}$ that
  \begin{itemize}
   \item  For  $\ell_{k_1}\ell_{\bar k_2}> 0$ is a $|\ell_{k_1}|\times |\ell_{\bar k_2}|$-block bi-infinite Hankel matrix.
   \item For $\ell_{k_1}\ell_{\bar k_2}< 0$ is a $|\ell_{k_1}|\times |\ell_{\bar k_2}|$-block bi-infinite Toeplitz matrix.
    \item For $\ell_{k_1}=0$ with $\ell_{\bar k_2}\neq 0$ we have a diagonal band structure being $|\ell_{\bar k_2}|$ its width, and for  $\ell_{\bar k_2}=0$ with $\ell_{ k_1}\neq 0$  a $|\ell_{k_1}|\times|\ell_{k_1}|$ block bi-infinite matrix.
  \end{itemize}
\end{pro}
\begin{proof}
  See Appendix B
\end{proof}
The Toeplitz/Hankel block structure appears not only  in the structure of $g_{k_1k_2}$ but also in the structure of $g$ itself, thought as an element
in $M_\Z(M_N(\C))$, for example if one takes $\ell_{k}=-\ell_{\bar k}=1$, $k=1,\dots,N$
we get a $N\times N$-block bi-infinite Toeplitz matrix, while for  $\ell_{k}=\ell_{\bar k}=1$, $k=1,\dots,N$ we get a $N\times N$-block bi-infinite Hankel matrix.

Notice that for the particular case $\ell_k=\ell_{\bar k}$, $k=1,\dots,N$, we have that $g$ is a block Hankel bi-infinite matrix and
 \begin{align*}
   g\sum_{k=1}^NE_{kk}\Lambda^{-\ell_k}&=\sum_{k=1}^NE_{kk}\Lambda^{\ell_k}g, &
   g\sum_{k=1}^NE_{kk}\Lambda^{\ell_k}&=\sum_{k=1}^NE_{kk}\Lambda^{-\ell_k}g.
 \end{align*}
 From these two equivalent conditions on $g$ we conclude for $g^2$ the following constraint
\begin{align*}
  g^2\sum_{k=1}^NE_{kk}\Lambda^{\ell_k}=\sum_{k=1}^NE_{kk}\Lambda^{\ell_k}g^2;
\end{align*}
i.e. $g^2$ is a block Toeplitz bi-infinite matrix and the corresponding solution to $g^2$  of periodic type with bared and non bared  periods equal to each other for each component: $\ell_{\bar k}=-\ell_k$, $k=1,\dots,N$.

In the one component case we get the condition
\begin{align}\label{1component constraint}
  L^{\ell_1}=\bar L^{-\ell_{\bar 1}}.
\end{align}
If $\ell_1+\ell_{\bar 1}=0$  we may choose $\ell_1=\ell\in\N$ and $\ell_{\bar 1}=-\ell$ and the constraint for $g$ is $g\Lambda^\ell=\Lambda^\ell g$ which leads $  L^{\ell}=\bar L^{\ell}$, i.e., the
$\ell$-th periodic reduction of the one component 2D Toda hierarchy \cite{ueno-takasaki}.
When $\ell_1,\ell_{\bar 1}>0$ are two nonnegative integers this constraint \eqref{1component constraint} gives the the reduction of the one component 2D Toda hierarchy suitable to be extended with additional flows as described in
\cite{carlet}, named there as bigraded. This is why we refer to this reduction when all $\ell_a$ are positive as multigraded reduction. Notice that this multigraded constraint over $g$ is never  of periodic type and  $\bS=\bS_+$ and $\bar\bS=\bar\bS_+$.

\paragraph{1D reduction and generalizations}
Given a set of nonnegative integers $\{\ell_a\}_{a\in\mathcal S}$ we request $g$ in \eqref{facW} the following constraint
\begin{align*}
  g\cdot  \Big(\sum_{k=1}^NE_{kk}\big(\Lambda^{\ell_{\bar k}}+\Lambda^{-\ell_{\bar k}}\big)\Big)
&=
  \Big(\sum_{k=1}^NE_{kk}\big(\Lambda^{\ell_k}+\Lambda^{-\ell_k}\big)\Big)\cdot g.
  \end{align*}
Now, as $z^j+z^{-j}=(z+z^{-1})^j+a_{j,j-2}(z+z^{-1})^{j-2}+\dots+a_{j,0}$, for some $a_{j,i}\in\Z$  we have
  \begin{align*}
 g\cdot \Big(\sum_{k=1}^NE_{kk}\big(\Lambda^{j\ell_{\bar k}}+\Lambda^{-j\ell_{\bar k}}\big)\Big)
 &=
\Big(\sum_{k=1}^NE_{kk}\big(\Lambda^{j\ell_k}+\Lambda^{-j\ell_k}\big)\Big) \cdot g
\end{align*}
and therefore
\begin{align*}
\sum_{k=1}^NC_{kk}\big(L^{j\ell_k}+L^{-j\ell_k}\big)=\sum_{k=1}^N\bar C_{kk}\big(\bar L^{j\ell_{\bar k}}+\bar L^{-j\ell_{\bar k}}\big)
\end{align*}
is fulfilled for any $j\geq 0$.

From here we conclude that
\begin{align*}
  \sum_{k=1}^N(B_{j\ell_k,k}+B_{j\ell_{\bar k},\bar k})=
  \sum_{k=1}^N C_{kk}\big(L^{j\ell_k}+L^{-j\ell_k}\big)=\sum_{k=1}^N\bar C_{kk}\big(\bar L^{j\ell_{\bar k}}+\bar L^{-j\ell_{\bar k}}\big)
\end{align*}
and therefore we deduce the invariance
\begin{align}\label{1D invariance}
  \sum_{k=1}^N(\partial_{j\ell_k,k}+\partial_{j\ell_{\bar k},\bar k})L=
  \sum_{k=1}^N(\partial_{j\ell_k,k}+\partial_{j\ell_{\bar k},\bar k})\bar L=0.
\end{align}
In the one component case if we choose $\ell_1=\ell_{\bar 1}=1$ we get the invariance under
$\partial_{j1}+\partial_{j\bar 1}$, $j>0$. This is the 1 dimensional reduction as discussed for example in \cite{ueno-takasaki}.

It must be stressed here that
being the same invariance conditions \eqref{1D invariance} for  this reduction and the previous multigraded reduction, the conditions are  different for $g$ and therefore for the class of solutions considered in the 2D Toda hierarchy.
In fact the Ueno--Takasaki 1D-reduction has soliton solutions, which appear as a particular class of the general
 soliton solutions of his  2D Toda hierarchy. However this Ueno--Takasaki's family of  soliton solutions of 2D Toda
  does not admit the bigraded type condition. On the other hand, the condition $L=\bar L^{-1}$ appears as a string equation in 2D Toda leading to solutions of the 1-matrix models, see for example \cite{takasaki string}.

\section{Orlov--Schulman operators, undressing,  and string equations}
\subsection{Introducing the Orlov--Schulman operator}
Given solutions $W,\bar W$ of the factorization problem \eqref{facW} we introduce the Orlov--Schulman operators \cite{orlov}
for the multicomponent 2D Toda hierarchy
 \begin{definition}
 The Orlov--Schulman operators are defined as follows
\begin{align}
  \label{orlov}
  M&:=Wn W^{-1},& \bar M&:=\bar Wn \bar W^{-1},
\end{align}
 \end{definition}
\begin{pro}\label{pro:orlov}
\begin{itemize}
\item The Orlov--Schulman operators satisfy the following commutation relations
\begin{gather}\label{algrebraic-orlov}
\begin{aligned}
{}[L,M]&=L,& [M,C_{kk}]&=0,&
[\bar L,\bar M]&=\bar L,& [\bar M,\bar C_{kk}]&=0,
\end{aligned}
\end{gather}
\item The following relations hold
\begin{gather}\label{orlov-exp}
\begin{aligned}
  M&=\m+
  \sum_{k=1}^NC_{kk}(s_k+\sum_{j=1}^\infty j t_{jk}L^j),& \m&=n+ \g_-\\
\bar M&=\bar \m-
  \sum_{k=1}^N\bar C_{kk}( s_{\bar k}+\sum_{j=1}^\infty j  t_{j\bar k}\bar L^{-j}),&\bar \m&=n+\g_+\Lambda.
\end{aligned}
\end{gather}
\end{itemize}
\end{pro}
\begin{proof}
 See Appendix B.
\end{proof}

Given the initial condition $g\in G$ in the factorization problem \eqref{facW} we write
\begin{align*}
g\Lambda E_{kk}g^{-1}&=\sum_{l,l'=1}^N p_{k,ll'}(n,\Lambda)E_{ll'},&
gnE_{kk} g^{-1}&=\sum_{l,l'=1}^N q_{k,ll'}(n,\Lambda)E_{ll'}
\end{align*}
and define
\begin{align}
\label{def:PQ}
\begin{aligned}
P_k&:=\sum_{l,l'=1}^N p_{k,ll'}(M,L)C_{ll'},&
Q_k&=\sum_{l,l'=1}^N q_{k,ll'}(M,L)C_{ll'},
\end{aligned}
\end{align}
so that
\begin{align}\label{pq}
  [P_k,Q_{k'}]=\delta_{kk'}P_k.
\end{align}
Then, as $Wg=\bar W$, we get, in the language of \cite{takasaki-takebe}, the  string  equations
\begin{gather}\label{twistor}
\begin{aligned}
 \sum_{l,l'=1}^N p_{k,ll'}(M,L)C_{ll'}&:=\bar L\bar C_{kk},&
\sum_{l,l'=1}^N q_{k,ll'}(M,L)C_{ll'}&:=\bar M\bar C_{kk}.
\end{aligned}
\end{gather}

\subsection{Undressing Lax equations for the Lax and Orlov--Schulman operators}

The Orlov--Schulman operators $M=WnW^{-1}$, $\bar M=\bar W n\bar W^{-1}$ satisfy
\begin{align*}
  \partial_{ja}M&=[\partial_{ja}W\cdot W^{-1},M],&   \partial_{ja}\bar M&=[\partial_{ja}\bar W\cdot\bar W^{-1},\bar M],\\
  T_KM&=(T_KW\cdot W^{-1})M(T_KW\cdot W^{-1})^{-1},&
    T_K\bar M&=(T_K\bar W\cdot\bar  W^{-1})\bar M(T_K\bar W\cdot\bar  W^{-1})^{-1}
\end{align*}
and  the factorization problem \eqref{facW} holds then the results of Theorem \ref{pro:integrable hierarchies}  imply the following Lax equations for the Orlov--Schulman operators
\begin{gather}\label{orlov-lax}
\begin{aligned}
  \partial_{ja} M&=[B_{ja},M],&  \partial_{ja}\bar M&=[B_{ja},\bar M],\\
  T_KM&=\omega_K M\omega_K^{-1},&
    T_K\bar M&=\omega_K \bar M\omega_K^{-1}.
\end{aligned}
\end{gather}
We now prove the local equivalence between the factorization problem and the Lax equations.

\begin{theorem}\label{undressing lax orlov}
Let us suppose that:
\begin{enumerate}
\item The operators $L,\bar
L,C_{kk},\bar C_{kk}, M$ and $\bar M$ satisfy the conditions
\begin{gather}\label{asymptotic}
\begin{aligned}
 L&=\Lambda+u_1(n)+u_2(n)\Lambda^{-1}+\cdots,&
\bar L^{-1}&=\bar u_0(n)\Lambda^{-1}+\bar u_1(n)+\bar u_2(n)\Lambda+\cdots,&  \\
C_{kk}&=E_{kk}+C_{kk,1}(n)\Lambda^{-1}+C_{kk,2}(n)\Lambda^{-2}+\cdots,&
\bar C_{kk}&=\bar C_{kk,0}(n)+\bar C_{kk,1}(n)\Lambda+\bar
C_{kk,2}(n)\Lambda^{2}+\cdots,\\
  M&=\cdots+M_{-1}\Lambda^{-1}+n  \sum_{k=1}^NC_{kk}(s_k+\sum_{j=1}^\infty j t_{jk}L^j),&
\bar M&=\cdots+\bar M_{1}\Lambda+n+
  \sum_{k=1}^N\bar C_{kk}( s_{\bar k}+\sum_{j=1}^\infty j  t_{j\bar k}\bar L^{-j}),
\end{aligned}
\end{gather}
with $k=1,\dots,N$, $\bar u_0(n)\in\operatorname{GL}(N,\C)$, and fulfill the equations
\begin{align}\label{eq:algebraic conditions}
\begin{aligned}
\mathbb{I}_N&=\sum_{k=1}^NC_{kk},&
C_{kk}C_{ll}&=\delta_{kl}C_{kk},& C_{kk}L&=LC_{kk},&C_{kk}M&=MC_{kk},&LM&=ML,\\
\mathbb{I}_N&=\sum_{k=1}^N\bar C_{kk},& \bar C_{kk}\bar
C_{ll}&=\delta_{kl}\bar C_{kk},& \bar C_{kk}\bar L&=\bar L\bar
C_{kk},&\bar C_{kk}\bar M&=\bar M\bar C_{kk},&\bar L\bar M&=\bar M\bar L.
\end{aligned}
\end{align}
\item Given operators $B$ and $\omega$ as in \eqref{omega}, the Lax equations
\eqref{laxtjk}, \eqref{dlaxtkl} and \eqref{orlov-lax} hold.
\end{enumerate}
Then, there exists operators $S\in G_-$ and $\bar S\in G_+$ such that for
 $W=SW_0$ and $\bar W=\bar S\bar W_0$ we may write
\begin{align*}
L&=W\Lambda W^{-1},& M&=Wn W^{-1},&C_{kk}&=WE_{kk}W^{-1},\\
\bar L&=\bar W\Lambda \bar W^{-1},&\bar M&=\bar W n\bar W^{-1},& \bar C_{kk}&=\bar WE_{kk}\bar W^{-1},
\end{align*}
 so that  $W$ and $\bar W$ solve the
factorization problem \eqref{facW} for some constant operator $g$.
\end{theorem}
Notice that the set of constrains \eqref{asymptotic} and  \eqref{eq:algebraic conditions} are preserved by the Lax equations.
\begin{proof}
Observe thatwe need  to find is the representation
\begin{align*}
L&=S\Lambda S^{-1},& M&=S\mu S^{-1},&C_{kk}&=SE_{kk}S^{-1},\\
\bar L&=\bar S\Lambda \bar S^{-1},&\bar M&=\bar S\bar\mu\bar S^{-1},& \bar C_{kk}&=\bar SE_{kk}\bar S^{-1},
\end{align*}
with $\mu$ and $\bar\mu$ as in \eqref{AdE}.
 We first undress the Lax operators $L,\bar L^{-1}$. We look for
\begin{align*}
S'&=\I_N+\varphi_1'(n)\Lambda^{-1}+\varphi_2'(n)\Lambda^{-2}+\cdots,& &\varphi'_j:\Z\to\gl,\\
\bar S'&=\bar \varphi_0'(n)(\I_N+\bar \varphi_1'(n)\Lambda+\bar \varphi_2'(n)\Lambda^{2}+\cdots),& &\bar \varphi_0':\Z\to\text{GL}(N,\C),\quad \bar \varphi'_j:\Z\to\gl,\; j>0
\end{align*}
such that
\begin{align}
\label{u1}(\Lambda+u_1+u_2\Lambda^{-1}+\cdots)(\I_N+\varphi'_1\Lambda^{-1}+\varphi'_2\Lambda^{-2}+\cdots)&=
(\I_N+\varphi'_1\Lambda^{-1}+\varphi_2'\Lambda^{-2}+\cdots)\Lambda,\\
\label{u2}(\bar u_0\Lambda^{-1}+\bar u_1+\bar u_2\Lambda^{2}+\cdots)\bar \varphi_0'(\I_N+\bar \varphi'_1\Lambda+\bar \varphi'_2\Lambda^2+\cdots)&=\bar \varphi_0
'(\I_N+\bar \varphi'_1\Lambda+\bar \varphi'_2\Lambda^2+\cdots)\Lambda^{-1}.
\end{align}
 Therefore,  if we define $\mathbb\Sigma_\pm [f]:=\sum_{j=0}^\infty f(n\pm j)$, we have
\[
\varphi'_1=c_1+\mathbb\Sigma_+[u_1].
\]
Once $\varphi'_1$ is obtained we fix our attention in the second
equation to get
\begin{align*}
\varphi'_2(n)&=c_2+\mathbb\Sigma_+[ u_2(n)+u_1 \varphi'_1].
\end{align*}
So forth and so on we get all the coefficients $\varphi'_j$ up to integration constants $c_j$.

Now we analyze \eqref{u2}, which we write as follows
\[
(\bar v_0\Lambda^{-1}+\bar v_1+\bar v_2\Lambda^{2}+\cdots)(\I_N+\bar \varphi'_1\Lambda+\bar \varphi'_2\Lambda^2+\cdots)=\Lambda^{-1}+\bar \varphi'_1+\bar \varphi'_2\Lambda+\cdots
\]
with $\bar v_j(n):=\bar \varphi_0'(n)^{-1}\bar u_j(n)\bar
\varphi_0'(n+j-1)$. From this we deduce that $\bar v_0=\I_N \text{ or
}\bar u_0(n)=\bar \varphi_0'(n)\bar \varphi_0'(n-1)^{-1}$. Denoting
$\phi':= \log\bar\varphi_0'$ we get $\log \bar u_0=(1-\Lambda^{-1})
(\phi')$ and therefore $\phi'(n)=\exp\big( \bar c_0+\mathbb\Sigma_-[
\log \bar u_0]\big)$ where $\bar c_0$ is a constant matrix. As $\bar
v_0=\I_N$ we have
\[
(\Lambda^{-1}+\bar v_1+\bar v_2\Lambda^{2}+\cdots)(\I_N+\bar \varphi'_1\Lambda+\bar \varphi'_2\Lambda^2+\cdots)=\Lambda^{-1}+\bar \varphi'_1+\bar \varphi'_2\Lambda+\cdots
\]
which transmutes into \eqref{u1} once we replace $u_j$ by $\bar v_j$, $\varphi_j'$ by $\bar\varphi_j'$, $j=1,2,\dots$ and $\Lambda$ by $\Lambda^{-1}$. Thus, all the coefficients $\bar \varphi_j'$ are expressed in terms of $\bar v_j$.

Now, we proceed to undress  $C_{kk}$ and $\bar C_{kk}$
\[
C'_{kk}:=S'^{-1}C_{kk} S',\quad \bar C'_{kk}:=\bar
S'^{-1}\bar C_{kk}\bar S'.
\]
These operators commute with $\Lambda$, satisfy
\begin{align*}
C_{kk}'&=E_{kk}+C'_{kk,1}\Lambda^{-1}+C'_{kk,2}\Lambda^{-2}+\cdots\\
\bar C'_{kk}&=\bar C'_{kk,0}+\bar C'_{kk,1}\Lambda+\bar
C'_{kk,2}\Lambda^{2}+\cdots,
\end{align*}
and provide us with two different resolutions of the identity,
 \begin{align*}
\mathbb{I}_N&=\sum_{k=1}^NC'_{kk},&
C'_{kk}C'_{ll}&=\delta_{kl}C'_{kk},&
\mathbb{I}_N&=\sum_{k=1}^N\bar C'_{kk},& \bar C'_{kk}\bar
C'_{ll}&=\delta_{kl}\bar C'_{k}.
\end{align*}
In fact, it is easy to show that there exists operators
$Q=\I_N+Q_1\Lambda^{-1}+\cdots\in G_-\cap \mathfrak
z_\Lambda$ and  $\bar Q=\bar Q_0+\bar Q_1\Lambda+\cdots\in
G_+\cap \mathfrak z_\Lambda$ such that $C_{kk}'=QE_{kk}Q^{-1}$
and $\bar C_{kk}'=\bar QE_{kk}\bar Q^{-1}$. Thus,   to undress
$L,\bar L,C_{kk},\bar C_{kk}$, $k=1,\dots , N$, we just take
$S=S'\cdot Q$, $\bar S=\bar S'\cdot \bar Q$.


 With these operators at hand we proceed to undress $M$ and $\bar M$
\begin{align*}
S^{-1}MS&=\alpha+\mu,& \alpha&:=S^{-1}\m S-n,&
\bar S^{-1}\bar M \bar S&=\bar\alpha+\bar\mu,& \bar\alpha&:=\bar S^{-1}\bar \m\bar S-n,
\end{align*}
but
\begin{align*}
[\Lambda,S^{-1}MS]&=\Lambda,& [E_{kk},S^{-1}MS]&=0,&
[\Lambda,\bar S^{-1}\bar M \bar S]&=\Lambda,& [E_{kk},\bar S^{-1}\bar M \bar S]&=0
\end{align*}
and $[\Lambda,\mu]=\Lambda$ and $[\Lambda,\bar\mu]=\Lambda$, so that
\begin{align*}
 [\Lambda,\alpha]&=0,& [E_{kk},\alpha]&=0,&
[\Lambda,\bar\alpha]&=0,& [E_{kk},\bar\alpha]&=0
& &\Rightarrow&\alpha,\bar\alpha&\in\h.
\end{align*}
Now, recalling that $\m=n+\g_-$ and $\bar\m=n +\Lambda \g_+$ we write
$\alpha=S^{-1}nS-n+\g_-$ and  $\bar\alpha=\bar S^{-1} n\bar S-n+ \g_+\Lambda$ so that $\alpha=\alpha_1\Lambda^{-1}+\alpha_2\Lambda^{-2}+\cdots$ and
$\bar\alpha=\bar\alpha_1\Lambda+\bar\alpha_2\Lambda^2+\cdots$ with $\alpha_i,\bar\alpha_i\in\diag$ for all $i\in\N$.
We define $\gamma:=-\sum_{j\geq 1}\frac{\alpha_j}{j}\Lambda^{-j}$,
$\bar\gamma:=\phi_0+\sum_{j\geq 1}\frac{\bar\alpha_j}{j}\Lambda^{j}$,
where $\phi_0\in\diag$,  and find that
\begin{align*}
\Exp{\gamma}n\Exp{-\gamma}&=n+[\gamma,n]=n+\alpha=S^{-1}\m S,&
\Exp{\bar\gamma}n\Exp{-\bar\gamma}&=n+[\bar\gamma,n]=n+\bar\alpha=\bar S^{-1}\bar\m \bar S
\end{align*}
which allows us to write
\begin{align*}
\Exp{\gamma}W_0nW_0^{-1}\Exp{-\gamma}&=\Exp{\gamma}n\Exp{-\gamma}+\nu=S^{-1}\m S+\nu,&
\Exp{\bar\gamma}\bar W_0n\bar W_0^{-1}\Exp{-\bar\gamma}&=\Exp{\bar\gamma}n\Exp{-\bar\gamma}+\bar\nu=\bar S^{-1}\bar\m \bar S+\bar\nu.
\end{align*}
Therefore, if we replace $S\to S\Exp{\gamma}$ and $\bar S\to \bar S\Exp{\bar\gamma}$,
we get the desired result.

From the evolution equation we get that
\begin{align}\label{eq:A-rho}
 \begin{aligned}
   A_{ja}&:=(S)^{-1}\cdot(B_{ja}-\partial_{ja}S\cdot (S)^{-1})S,& \bar A_{ja}&:=(\bar S)^{-1}\cdot( B_{ja}-\partial_{ja}\bar
  S\cdot (\bar S)^{-1})\bar S,\\
\rho_K&:= T_K(S)^{-1}\omega_K S,& \bar\rho_K&:= T_K(\bar
S)^{-1}\omega_K \bar S,
 \end{aligned}
\end{align}
 commute with $\Lambda$ and all the $E_{kk}$, $k=1,\dots,N$; i.e., they are
$n$-independent and diagonal.
We  may deduce that
\begin{gather}\label{eq:A-theta}
\begin{aligned}
 {} [A_{jn},\mu]&=j\theta_{ja}\Rightarrow [A_{ja},n]=j\theta_{ja}\Rightarrow [A_{ja}-\theta_{ja},n]=0,\\
  \\
  [\bar A_{jn},\bar \mu]&=-j\bar\theta_{ja}\Rightarrow
   [\bar A_{ja},n]=-j\bar \theta_{ja}\Rightarrow [\bar A_{ja}-\bar \theta_{ja},n]=0,
\end{aligned}
\end{gather}
and
\begin{align*}
  \omega_KM\omega_K^{-1}&= (T_KS\cdot S^{-1})M(T_KS\cdot S^{-1})^{-1}+(T_KS)(\pi_a-\pi_b) (T_KS)^{-1},\\
    \bar\omega_K\bar M\bar\omega_K^{-1}&= (T_K\bar S\cdot \bar S^{-1})
    \bar M(T_K\bar S\cdot S^{-1})^{-1}-(T_K\bar S)(\bar \pi_a-\bar \pi_b) (T_K\bar S)^{-1},
\end{align*}
which imply
\begin{gather}\label{eq:rho-q}
\begin{aligned}
  {}[\rho_k,\mu]&=(\pi_a-\pi_b)\rho_k\Rightarrow  [\rho_k,n]=(\pi_a-\pi_b)\rho_k\Rightarrow[\rho_kq_K^{-1},n]=0,\\
   [\bar\rho_k,\bar\mu]&=-(\bar\pi_a-\bar\pi_b)\bar\rho_k\Rightarrow [\bar\rho_k,n]=
   -(\bar\pi_a-\bar\pi_b)\bar\rho_k\Rightarrow [\bar\rho_k\bar q_K^{-1},n]=0.
\end{aligned}
\end{gather}
Thus,
\begin{align}\label{a-rho-diag}
A_{ja}-\theta_{ja},\,\bar A_{ja}-\bar \theta_{ja}, \,\rho_kq_K^{-1},\,\bar\rho_k\bar q_K^{-1}\in\diag.
\end{align}

As the Lax equations \eqref{laxtjk} and \eqref{dlaxtkl} are satisfied by Proposition \ref{lax then zs}
 we know that $B_{ja}$ and $\omega_K$ satisfy the compatibility
conditions \eqref{zs}-\eqref{con-omega}. However, we see from \eqref{eq:A-rho} that  $\{A_{ja},\rho_K\}$ and
$\{\bar A_{ja},\bar\rho_K\}$ are gauge transforms of $B_{ja},\omega_K$ and thereby do have zero curvature.
Therefore, we conclude the
local existence of potentials $\xi$ and $\bar\xi$ such that
\begin{align}
\label{eq:xi}
\begin{aligned}
   A_{ja}&=\partial_{ja}\xi\cdot\xi^{-1},&\bar
  A_{ja}&=\partial_{ja}\bar\xi\cdot\bar\xi^{-1},&
  \rho_K&=T_K\xi\cdot\xi^{-1},&
  \bar\rho_K&=T_K\bar\xi\cdot\bar\xi^{-1}.
\end{aligned}
\end{align}
These potentials are determined up to right multiplication $ \xi\to
\xi\cdot h,\quad \bar\xi\to\bar\xi\cdot \bar h$, where $h,\bar h\in
G$ are constant operators independent of $t_{ja},s_a$. Up to this
freedom we may take the potentials $\xi,\bar\xi\in H$.
 Now, recalling \eqref{omega} we get
\begin{align*}
B_{ja}&=\r_{ja}+\g_-=\bar\r_{ja}+\g_+& \omega_K,&=G_-\cdot \u_K=G_+\cdot \bar \u_K,
\end{align*}
 which  together with \eqref{eq:A-rho} imply $A_{ja}-\theta_{ja}\in\g_-$,
 $\bar A_{ja}-\bar\theta_{ja}\in\g_+$, $\rho_Kq_K^{-1}\in G_-$, $\bar \rho_K\bar q_K^{-1}\in G_+$;
but these operators belong to $\diag$. Thus, from the two first relations we conclude
$ A_{ja}=\theta_{ja}$, $\rho_K=q_K$  and $\xi=W_0$
while the two second imply that if
$\bar\xi=\Exp{\phi_0}\cdot \bar W_0$ then $\phi_0\in\diag$.

 Therefore,  we may write
\begin{align*}
  A_{ja}&=\partial_{ja}W_0\cdot W_0^{-1},& \bar A_{ja}&=
  \partial_{ja} (\bar\xi\bar W_0)\cdot (\Exp{\phi_0}\bar W_0)^{-1},&
  \rho_K&=T_KW_0\cdot W_0^{-1},& \bar\rho_K&=T_K(\Exp{\phi_0}\bar W_0)\cdot (\Exp{\phi_0}\bar W_0)^{-1}.
\end{align*}
 We make the replacement $\bar S\to \bar S\Exp{\phi_0}$ to get
\begin{align*}
  B_{ja}=\partial_{ja}W\cdot W^{-1}=\partial_{ja}S\cdot S^{-1}+S\theta_{ja}S^{-1}&=\partial_{ja}\bar S\cdot\bar S^{-1}+\bar S\bar\theta_{ja}\bar S^{-1}=\partial_{ja}\bar W\cdot \bar W^{-1},\\
  \omega_K=(T_KW)\cdot W^{-1}=(T_KS) q_k S^{-1}&=(T_K\bar S) \bar q_K\bar S^{-1}=(T_K\bar W)\cdot\bar W^{-1}.
\end{align*}
In terms of $g= W^{-1}\cdot \bar W$ the previous equations can be written as $\partial_{ja}g=0$ and $T_Kg=g$.
 Thus, we finally find  $W g=\bar W$ where $g$ is a constant operator in $G$.
\end{proof}


A further result regarding the operators $C_{kl}$ introduced in Definition \ref{lax-C} and characterized in Proposition \ref{lax-C2} that will be needed later is given now.

\begin{pro}\label{undressing Ckl}
  Given operators $L,\bar L,M,\bar M,C_{kk}$ and $\bar C_{kk}$ as in Theorem \ref{undressing lax orlov}, then:
     if we find operators $C_{kl}$ of the form
    \begin{align*}
    C_{kl}&=L^{s_k-s_l}\Exp{(t_{jk}-t_{jl})L^j}(E_{kl}+\g_-),
  \end{align*}
  such that
  \begin{align*}
    &[C_{kl},L]=[C_{kl},M]=0,&
      &C_{k'k'}C_{kl}=\delta_{kk'}C_{k'l},\quad C_{kl}C_{k'k'}=\delta_{kl'}C_{kk'},
  \end{align*}
  then  the undressing operator $W$  of Theorem \ref{undressing lax orlov} satisfies
$ C_{kl}=W E_{kl}W^{-1}$.
\end{pro}
\begin{proof}
  See Appendix B.
\end{proof}

\subsection{String equations, factorization problem and  Lax equations }

 We will show here that the string equations \eqref{twistor} for the Lax and Orlov--Schulman operators do indeed imply the factorization \eqref{facW} and also the Lax equations \eqref{laxtjk}-\eqref{dlaxtkl}. In fact, only one of these implications is needed as the other one will follow from the results described previously.
 However, we show that these two facts can be derived directly from the string  equations,
 showing the importance of this formulation of integrable systems.

\begin{theorem}
  \label{undresing twistor}
  Let $L,M,C_{kk},\bar L,\bar M,\bar C_{kk}$, $k=1,\dots,N$, be operators as in Theorem
   \ref{undressing lax orlov} and operators $C_{kl}$, $k,l=1,\dots,N$,  be as in Proposition \ref{undressing Ckl}. Let us suppose that we have operators $P_k,Q_k$ as in \eqref{def:PQ} and that the string equations \eqref{twistor} hold. Then,
  \begin{enumerate}
  \item We can choose the operators $W$ and $\bar W$ of Theorem  \ref{undressing lax orlov}
  such that the factorization \eqref{facW} holds for some constant operator $g\in G$.
    \item  The Lax equations \eqref{laxtjk}-\eqref{dlaxtkl} are fulfilled.
  \end{enumerate}

\end{theorem}
\begin{proof}
  From the first part of the proof of Theorem \ref{undressing lax orlov} (not considering the Lax equations) and Proposition \ref{undressing Ckl} we know that there are undressing operators $W=SW_0$ and $\bar W=\bar S\bar W_0$, $S\in G_-$ and $\bar S\in G_+$. Let us introduce some convenient notation
  \begin{align}\label{Dsigma}
  \begin{aligned}
        D_{ja}&:=\partial_{ja}W\cdot W^{-1}-\partial_{ja}\bar W\cdot\bar W^{-1}&   D_{ja}^0&:=\bar W^{-1}D_{ja}\bar W,\\
        \sigma_K&:=(T_K\bar W\cdot\bar W^{-1})^{-1}T_KW\cdot W^{-1}& \sigma_K^0&:=\bar W^{-1}\sigma_K\bar W,
      \end{aligned}
      \end{align}
      and observe that if we define
      \begin{align}\label{zeta}
        \zeta:=\bar W^{-1}\cdot W
      \end{align}
      we have
      \begin{align}\label{zetadsigma}
        D_{ja}^0&=\partial_{ja}\zeta\cdot\zeta^{-1},&\sigma_K^0=T_K\zeta\cdot\zeta^{-1}.
      \end{align}
The string equations \eqref{twistor} read
\begin{align}\label{undressing pq}  \begin{aligned}
    P_k&=WP_k^0W^{-1}=\bar W E_{kk}\Lambda \bar W^{-1},& P_k^0&=\sum_{l,l'=1}^N p_{k,ll'}(n,\Lambda)E_{ll'},\\
    Q_k&=WQ_k^0W^{-1}=\bar W E_{kk}n \bar W^{-1},& Q_k^0&=\sum_{l,l'=1}^N q_{k,ll'}(n,\Lambda)E_{ll'},\\
      \end{aligned}
      \end{align}
      which in turn imply
      \begin{align*}
        \partial_{ja}P_k&=[\partial_{ja}W\cdot W^{-1},P_k]=[\partial_{ja}\bar W\cdot \bar W^{-1},P_k],\\
        \partial_{ja}Q_k&=[\partial_{ja}W\cdot W^{-1},Q_k]=[\partial_{ja}\bar W\cdot \bar W^{-1},Q_k],\\
        T_KP_k&=(T_KW\cdot W^{-1})P_k(T_KW\cdot W^{-1})^{-1}=(T_K\bar W\cdot \bar W^{-1})P_k(T_K\bar W\cdot \bar W^{-1})^{-1},\\
        T_KQ_k&=(T_KW\cdot W^{-1})Q_k(T_KW\cdot W^{-1})^{-1}=(T_K\bar W\cdot \bar W^{-1})Q_k(T_K\bar W\cdot \bar W^{-1})^{-1}.
      \end{align*}
      Hence, recalling  $P_k=\bar L \bar C_{kk}$ and $Q_k=\bar M \bar C_{kk}$ we conclude that
      \begin{align}\label{dsigma}\begin{aligned}
       {} [D_{ja},\bar L]&=[D_{ja},\bar M]=[D_{ja},\bar C_{kk}]=0,&
         [\sigma_K,\bar L]&=[\sigma_K,\bar M]=[\sigma_K,\bar C_{kk}]=0
      \end{aligned}
      \end{align}
      Thus, we deduce that $ D_{ja}^0,\sigma_K^0\in\diag$,
      and therefore $D_{ja}\in\g_+$ and $\sigma_K\in G_+$.
    With these preliminaries let us start proving the two statements in the theorem  \begin{enumerate}
        \item Given the representation \eqref{zetadsigma} in terms of $\zeta$ as in \eqref{zeta} we deduce that we can write
           $\zeta=\bar\xi\cdot g^{-1}$ for some $\bar\xi\in \diag$ and some $(\bs,\bt)$- independent operator $g\in G$. Thus,
          $D_{ja}^0=\partial_{ja}\bar\xi\cdot\bar\xi^{-1}$ and $\sigma_K^0=T_K\bar\xi\cdot    \bar\xi^{-1}$.
                But, recalling the definition \eqref{zeta} we get $\bar W\zeta=W\Rightarrow \bar S\bar W_0\bar\xi=W g$.
               Observe that $[\bar W_0,\bar\xi]=0$ and replace
                $\bar S\rightarrow\bar S\cdot\bar\xi$ to get the factorization problem.
                      \item From the definition \eqref{Dsigma} we get
\begin{align*}
  \partial_{ja}W\cdot W^{-1}&=\partial_{ja}S\cdot S^{-1}+\r_{ja}=
  \partial_{ja}\bar S\cdot\bar S^{-1}+\bar\r_{ja}+D_{ja}=\partial_{ja}\bar W\cdot \bar W^{-1}+D_{ja},\\
  T_KW\cdot W^{-1} &=T_K S\cdot S^{-1} \cdot\u_K=T_K \bar S\cdot \bar S^{-1} \cdot\bar\u_K\cdot\sigma_K=T_K\bar W\cdot\bar W^{-1}\cdot\sigma_K.
\end{align*}
Reasoning as in the proof of Theorem \ref{undressing lax orlov} and recalling that $D_{ja}\in\g_+$,  $\sigma_K\in G_+$ and $[\bar\u_K,\sigma_K]=0$ we have
\begin{align*}
 \partial_{ja}S\cdot S^{-1}&=- (\r_{ja}-\bar\r_{ja})_-, & \partial_{ja}\bar S\cdot \bar S^{-1}+D_{ja}&= (\r_{ja}-\bar\r_{ja})_+,\\
 T_KS\cdot S^{-1} &=(\u_K\cdot\bar\u_K^{-1})_-,&T_K\bar S\cdot \bar S^{-1}\cdot \sigma_K &=(\u_K\cdot\bar\u_K^{-1})_+
\end{align*}
so that, according to \eqref{omega},
\begin{align*}
  \partial_{ja}W\cdot W^{-1}&=B_{ja}=\partial_{ja}\bar W\cdot \bar W^{-1}+D_{ja},&
  T_KW\cdot W^{-1} &=\omega_K=T_K\bar W\cdot\bar W^{-1}\cdot\sigma_K.
\end{align*}
Therefore, we immediately get the following Lax equations
\begin{align*}
  \partial_{ja} L&=[\partial_{ja}W\cdot W^{-1},L]=[B_{ja},L],& T_KL&=(T_KW\cdot W^{-1})L(T_KW\cdot W^{-1})^{-1}=\omega_KL\omega_K^{-1}\\
    \partial_{ja} M&=[\partial_{ja}W\cdot W^{-1},M]=[B_{ja},M],&T_KM&=(T_KW\cdot W^{-1})M(T_KW\cdot W^{-1})^{-1}=\omega_KM\omega_K^{-1}\\
      \partial_{ja} C_{kk}&=[\partial_{ja}W\cdot W^{-1},C_{kk}]=[B_{ja},C_{kk}],&T_KC_{kk}&=(T_KW\cdot W^{-1})C_{kk}(T_KW\cdot W^{-1})^{-1}=\omega_KC_{kk}\omega_K^{-1}.
\end{align*}
Now, as $\partial_{ja}\bar W\cdot \bar W^{-1}=B_{ja}-D_{ja}$ and
$ T_K\bar W\cdot \bar W^{-1}=\omega_K\cdot\sigma_K$
 with $D_{ja}$ and $\sigma_K$ commuting with any function of $\bar L$, $\bar M$ and $\bar C_{kk}$ we get the remaining Lax equations
\begin{align*}
  \partial_{ja} \bar L&=[\partial_{ja}\bar W\cdot \bar W^{-1},\bar L]=[B_{ja},\bar L],& T_K\bar L&=(T_K\bar W\cdot\bar W^{-1})\bar L(T_K\bar W\cdot \bar W^{-1})^{-1}=\omega_K\bar L\omega_K^{-1}\\
    \partial_{ja} \bar M&=[\partial_{ja}\bar W\cdot \bar W^{-1},\bar M]=[B_{ja},M],&T_K\bar M&=(T_K\bar W\cdot \bar W^{-1})\bar M(T_K\bar W\cdot \bar W^{-1})^{-1}=\omega_K\bar M\omega_K^{-1}\\
      \partial_{ja} \bar C_{kk}&=[\partial_{ja}\bar W\cdot \bar W^{-1},\bar C_{kk}]=[B_{ja},C_{kk}],&T_K\bar C_{kk}&=(T_K\bar W\cdot W^{-1})\bar C_{kk}(T_K\bar W\cdot \bar W^{-1})^{-1}=\omega_K\bar C_{kk}\omega_K^{-1}.
\end{align*}
      \end{enumerate}
\end{proof}
The above result might be slightly generalized by considering string equations of the form
 \begin{align*}
   \sum_{l,l'=1}^Np_{k,ll'}(M,L)C_{ll'}&=\sum_{l,l'=1}^N\bar p_{k,ll'}(\bar M,\bar L)\bar C_{ll'},&
   \sum_{l,l'=1}^Nq_{k,ll'}(M,L)C_{ll'}&=\sum_{l,l'=1}^N\bar q_{k,ll'}(\bar M,\bar L)\bar C_{ll'},
 \end{align*}
 where we assume that
 \begin{align*}
   \bar P_k^0&:=\sum_{l,l'=1}^N\bar p_{k,ll'}(n, \Lambda)\bar E_{ll'},& \bar Q_k^0&:=\sum_{l,l'=1}^N\bar q_{k,ll'}(n,\Lambda)\bar E_{ll'},\quad k=1,\dots,N,
 \end{align*}
 belong to the adjoint orbit $\mathscr O $ of $E_{kk}\Lambda$, $E_{kk}n$, $k=1,\dots,N$;
 i.e., there exists $\bar g\in G$ such that
 \begin{align*}
   P_k^0&=\bar g\cdot E_{kk}\Lambda\cdot\bar g^{-1}, &    Q_k^0&=\bar g\cdot E_{kk}n\cdot\bar g^{-1} .
 \end{align*}
For that aim the proof needs to be modified only in the definition of $D_{ja}^0\rightarrow\bar g^{-1}D_{ja}^0\bar g^{-1}$ and $\sigma_K^0\rightarrow\bar g^{-1}\sigma_K^0\bar g$ and $g\rightarrow g\cdot\bar g$. Observe that elements in $\mathscr O$ can be constructed in terms of operators $\mathcal C_k,\mathcal P$ and $\mathcal Q$, such that:
 \begin{align*}
   \sum_{k=1}^N\mathcal C_k&=\I_N, &
   \mathcal C_k\mathcal C_{k'}&=\delta_{kk'}\mathcal C_k,&
   [\mathcal C_k,\mathcal P]&=[\mathcal C_k,\mathcal Q]=0,&\quad [\mathcal P,\mathcal Q]&=\mathcal P.
 \end{align*}
 What we do not know yet is if the orbit $\mathscr O$ is characterized precisely by this properties.
  However, if we request the following properties: $\mathcal C_k-E_{kk}\in \g_-$,
  $\mathcal P-\Lambda\in \g_-\Lambda$, $\mathcal Q-n\in\g_-$,
 one could prove, following similar arguments as in the proof of Theorem \ref{undressing lax orlov}, that these elements lay in $\mathscr O$.  This implies an alternative formulation of string equations \eqref{twistor}
\begin{align*}
  \sum_{l,l'}\mathcal C_{k,ll'}(L,M)C_{ll'}&=\bar C_{kk},&
   \sum_{l,l'}\mathcal P_{ll'}(L,M)C_{ll'}&=\bar L,&
   \sum_{l,l'}\mathcal Q_{ll'}(L,M)C_{ll'}&=\bar M.
\end{align*}

\subsection{Additional symmetries and string equations}

\subsubsection{Additional symmetries}Suppose that the operator $g$ in \eqref{facW} depends
on an additional, or external, parameter $b$, which might belong to $\C$ or to $\Z$. Now, we
 will describe the induced dependence on the elements defining the multicomponent Toda hierarchy.
   We shall denote by $\partial_{\tb}=\partial/\partial \tb$ when $\tb\in \C$ is a continuous an by $T_{\tb}$
   the corresponding shift $\tb\rightarrow \tb+1$, when $\tb\in\Z$ is an integer.
In this case, we shall replace \eqref{facW} by the equivalent factorization problem
$  W\cdot h=\bar W\cdot\bar h$,
with
\begin{align}\label{factor g}
g=h\cdot\bar h^{-1}.
\end{align}
Observe that for $\tb\in\C$ we may write,
\begin{align}\label{external parameter continous}
\begin{aligned}
\partial_{\tb}W\cdot W^{-1}+W(\partial _{\tb}h\cdot h^{-1})W^{-1}&=\partial_{\tb} S\cdot S^{-1}+W(\partial _{\tb}h\cdot h^{-1})W^{-1}\\&=\partial_{\tb}\bar S\cdot\bar S^{-1}+\bar W(\partial _{\tb}\bar h\cdot\bar h^{-1})\bar W^{-1}=
\partial_{\tb}\bar W\cdot \bar W^{-1}+\bar W(\partial _{\tb}\bar h\cdot\bar h^{-1})\bar W^{-1},
\end{aligned}
\end{align}
while for $\tb\in \Z$ we have
\begin{align}\label{external parameter discrete}
\begin{aligned}
T_{\tb}W\cdot W^{-1}\cdot W\cdot(T_{\tb}h\cdot h^{-1})\cdot W^{-1}&=T_{\tb} S \cdot S^{-1}\cdot W\cdot(T_{\tb}h\cdot h^{-1})\cdot W^{-1}\\&=T_{\tb}\bar S\cdot\bar S^{-1} \cdot\bar W\cdot(T_{\tb}\bar h\cdot\bar h^{-1})\cdot \bar W^{-1}=
T_{\tb}\bar W\cdot \bar W^{-1} \cdot\bar W\cdot(T_{\tb}\bar h\cdot\bar h^{-1})\cdot \bar W^{-1}
\end{aligned}
\end{align}

Now we suppose that the dependence on $\tb$ is given by the following equations
\begin{align}\label{derivada g}
  \begin{aligned}
    \partial_{\tb}h\cdot h^{-1}&=F_0=\sum_{l,l'=1}^N F_{ll'}(n,\Lambda)E_{ll'},  &
     \partial_{\tb}\bar h\cdot \bar h^{-1}&=\bar F_0=\sum_{l,l'=1}^N \bar F_{ll'}(n,\Lambda)E_{ll'}& \text{ when } \tb\in\C,\\
     T_{\tb}h\cdot h^{-1}&=\mathcal F_0=\sum_{l,l'=1}^N  \mathcal F_{ll'}(n,\Lambda)E_{ll'},&
        T_{\tb}\bar h\cdot \bar h^{-1}&=\bar{\mathcal F}_0=\sum_{l,l'=1}^N  \bar{\mathcal F}_{ll'}(n,\Lambda)E_{ll'},
     ,& \text{ when } \tb\in\Z,
  \end{aligned}
\end{align}
and define
\begin{align}\label{fbarf}
  \begin{aligned}
    F&:=\sum_{l,l'=1}^N F_{ll'}(M,L)C_{ll'},&
     \bar F&:=\sum_{l,l'=1}^N \bar F_{ll'}(\bar M,\bar L)\bar C_{ll'}& \text{ when } \tb\in\C,\\
    \mathcal F&:=\sum_{l,l'=1}^N  \mathcal F_{ll'}(M,L)C_{ll'},&   \bar{ \mathcal F}&:=\sum_{l,l'=1}^N \bar{ \mathcal F}_{ll'}(\bar M,\bar L)\bar C_{ll'},& \text{ when }\tb\in\C.
  \end{aligned}
\end{align}
From \eqref{external parameter continous}-\eqref{external parameter
discrete} we get
\begin{align*}
   \partial_{\tb} W\cdot W ^{-1}= \partial_{\tb} S\cdot S^{-1}&=-(F-\bar F)_-, &
   \partial_{\tb} \bar W\cdot \bar W^{-1}=\partial_{\tb} \bar S\cdot \bar S^{-1}&=(F-\bar F)_+,  &(F-\bar F)_\pm&\in\g_\pm,\\
    T_{\tb} W\cdot W^{-1}=T_{\tb} S\cdot S^{-1}&=(\mathcal F\cdot \bar{\mathcal F}^{-1})_-, & T_{\tb} \bar W\cdot \bar W^{-1}=T_{\tb} \bar S\cdot \bar S^{-1}&=(\mathcal F\cdot \bar{\mathcal F}^{-1})_+& (\mathcal F\cdot \bar{\mathcal F}^{-1})_\pm&\in G_\pm.
\end{align*}
So that
\begin{pro}\label{pro:external parameters}
Given a dependence on an additional parameter $\tb$ according to \eqref{factor g} and \eqref{derivada g}, introduce $  H:=F-\bar F$ and $ \mathcal H:=\mathcal F\cdot \bar{\mathcal F}^{-1}$ where $F$ and $\bar F$ are defined \eqref{fbarf}, then
\begin{enumerate}
  \item The dressing operators $W$ and $\bar W$ satisfy the following equations
  \begin{align*}
    \partial_{\tb} W&=-H_-\cdot W, &\partial_{\tb}\bar W&=H_+\cdot\bar W,&
 &\text{or} &    T_{\tb} W &= \mathcal H_-\cdot W,& T_{\tb}\bar W&= \mathcal H_+\cdot\bar W.
 \end{align*}
\item The Lax and Orlov--Schulman operators are subject to
\begin{align}\label{additional flows}
\begin{aligned}
   \partial_{\tb}  L&=[-H_-,L],& \partial_{\tb} M&=[-H_-,M],& \partial_{\tb}  C_{kk}&=-[H_-,C_{kk}],\\
  \partial_{\tb}  \bar L&=[H_+,\bar L],& \partial_{\tb}  \bar M&=[H_+,\bar M],& \partial_{\tb} \bar C_{kk}&=[H_+,\bar C_{kk}],\\ &&&\text{or}&&\\
   T_{\tb} L&=\mathcal H_-\cdot L\cdot \mathcal H_-^{-1},&T_{\tb}  M&=\mathcal H_-\cdot M\cdot \mathcal H_-^{-1},&T_{\tb}  C_{kk}&=\mathcal H_-\cdot C_{kk}\cdot \mathcal H_-^{-1},\\
  T_{\tb}  \bar L&= \mathcal H_+\cdot\bar L\cdot  \mathcal H_+^{-1},&T_{\tb}  \bar M&= \mathcal H_+\cdot \bar M\cdot  \mathcal H_+^{-1},&T_{\tb}  \bar C_{kk}&=\mathcal H_+\cdot \bar C_{kk}\cdot \mathcal H_+^{-1}.
  \end{aligned}
\end{align}
\end{enumerate}
\end{pro}
\subsubsection{String equations}
The factorization problem \eqref{facW} depends decisively on the `initial data' $g$. Now, we are going to see some consequences of the form of $g$ and derive the so called string equations. Let us suppose, that given  two operators
\begin{align*}
  F_0&:=\sum_{l,l'=1}^N F_{ll'}(n,\Lambda)E_{ll'},& \bar F_0&=\sum_{l,l'=1}^N \bar F_{ll'}(n,\Lambda)E_{ll'},
\end{align*}
we have the following constraint satisfied by $g$
\begin{align}\label{prestring}
  g\bar F_0=F_0 g.
\end{align}
Then, if
\begin{align*}
  F(M,L)&:=\sum_{l,l'=1}^N F_{ll'}(M,L)C_{ll'},& \bar F(\bar M,\bar L)&=\sum_{l,l'=1}^N \bar F_{ll'}(\bar M,\bar L)\bar C_{ll'},
\end{align*}
we have
\begin{align}\label{string}
  F(M,L)=\bar F(\bar M,\bar L).
\end{align}

We refer to these type of equations as string equations, see for example \cite{takasaki string}, and we
have seen that they reflect properties like \eqref{prestring} of the initial condition $g$ in \eqref{facW}.
Notice that the reduction of \eqref{periodic g} is a particular case of \eqref{prestring} with $\bar F_0:=\sum_{k=1}^NE_{kk}\Lambda^{-\ell_{\bar k}}$ and $F_0:=\sum_{k=1}^NE_{kk}\Lambda^{\ell_{ k}}$,  in which the Orlov--Schulman operator does not appear.
This suggests an important family of diagonal string equations with
\begin{align}\label{diagonal string}
  F_0&:=\sum_{k=1}^NE_{kk}F_{0,k}(n,\Lambda), &\bar F_0&:=\sum_{k=1}^NE_{kk}\bar F_{0,k}(n,\Lambda).
\end{align}

The  equations \eqref{twistor} are also a set of string equations; moreover,
 the invariance conditions under the additional flow \eqref{additional flows} implies
 that $H=0$ or $\mathcal H=\text{id}$ so that we are lead to the string type equations
 of the form \eqref{string}, namely
\begin{align}\label{invariance conditions}
\begin{aligned}
   F(M,L)&=\bar F(\bar M,\bar L)& &\text{ or }&
   \mathcal F(M,L)&=\bar{\mathcal  F}(\bar M,\bar L).
\end{aligned}
\end{align}
This also follows from
\begin{align*}
    \partial_{\tb}g&=(\partial_{\tb} h\cdot h^{-1})g-g( \partial_{\tb} \bar h\cdot\bar h^{-1})=
    F_0 g-g\bar F_0,\\
    T_{\tb}g&= (T_{\tb}h\cdot h^{-1})\cdot g\cdot (T_{\tb}\bar h\cdot\bar h^{-1})^{-1}=\mathcal F_0\cdot g\cdot\bar{\mathcal F}_0^{-1}.
\end{align*}
Observe if we consider arbitrary forms of $F_0$, $\bar F_0$ or $\mathcal F_0$, $\bar{\mathcal F}_0$ it will be same to deal with the description given here or the one obtained just setting $\bar h=\text{id}$. However the situation is different if we consider the function $F_0$, $\bar F_0$ or $\mathcal F_0$, $\bar{\mathcal F}_0$ of diagonal type \eqref{diagonal string}. In this case, to set $\bar h=\text{id}$  will  generically imply to abandon the diagonal family for $F_0$.


\appendix

\section*{Appendix A: Congruences}
We will show here how to derive from the multi-component Toda
hierarchy  equations involving only
 the fields at each site $n\in\Z$; i.e. not mixing fields at different values of $n$, the sequence variable.
 This is particulary useful to show the role of the discrete multicomponent KP hierarchy in the multicomponent Toda hierarchy, which appears when we froze the bared continuous and discrete times.

 Firstly, we present a queue observation
\begin{pro}\label{gd}
Let us suppose that we have operators $R, \bar R\in\g$ such that
\begin{align}\label{nuevolemma}
\begin{aligned}
  RW_0^{-1}&\in\g_-,&
  \bar R\bar W_0^{-1}&\in\g_+
\end{aligned}
\end{align}
satisfying $ R\cdot g=\bar R$. Then $ R=\bar R=0$.
 \end{pro}
\begin{proof}
We have
\begin{align*}
 \bar R= Rg=RW^{-1}Wg=RW^{-1}\bar W\Rightarrow
 \bar R\bar W^{-1}=RW^{-1}
\end{align*}
therefore $  \bar R\bar W_0^{-1}\bar S^{-1}=RW_0^{-1}S^{-1}$ and
recalling \eqref{nuevolemma}, and the fact that $S\in G_-$ and $\bar
S\in G_+$ we conclude the statement.
\end{proof}

Next, and without proof (which consist in a systematic and sometimes
elaborated application of the previous result) we show the appearance
of some well known integrable hierarchies within the multicomponent
Toda hierarchy.   We firstly point out  that continuous variables, and for
each value of $n$, we have solutions of the $N$-wave hierarchy and its
modifications,  moreover some discretizations of the modified
$N$-wave equations are proposed. These results are just a
manifestation of the fact that if we froze the bared times we are just
dealing with a discrete $N$-component KP hierarchy in the spirit of
\cite{adler-van moerbeke}. Next, we recover within this context the the
quadrilateral lattice equations. Finally, we present what we call the
dispersive Whitham hierarchy in complete analogy to the one proposed
in \cite{misgam}-\cite{takasaki-takebe-ultimo}.

\paragraph{$N$-resonant wave equations and its modification}
We introduce
\begin{align*}
\partial&:=\partial_{11}+\dots+\partial_{1N},&
\bar\partial&:=\partial_{1\bar 1}+\dots+\partial_{1\bar N},
\end{align*}
in terms of which we  have
\begin{theorem}\label{pro-baker}
The dressing operators satisfy the following equations
\begin{align}\label{bakerevol1}
\partial_{ja}W&=Q_{ja}(W),&\partial_{ja}\bar W&=Q_{ja}(\bar W),
\end{align}
where
\begin{align*}
Q_{jk}&=u_{jk,j}\partial^j+u_{jk,j-1}\partial^{j-1}+\dots+u_{jk,0}, &
 Q_{j\bar k}&=v_{jk,j}\bar\partial^j+v_{jk,j-1}\bar\partial^{j-1}+\dots+v_{jk,1}\bar\partial
\end{align*}
with the coefficients $u_{jk,i}, v_{jk,i}$ depending on
$\partial_{1}^s\varphi_r, \bar\partial_{1}^s\bar\varphi_r$ ,
respectively
  \begin{align*}
u_{jk,j-i}&=\begin{cases}
E_{kk},& i=0,\\
\varphi_i E_{kk}-\sum_{a=0}^{i-1}u_{jk,j-a}\sigma_{j-a,i-a},& i=1,\dots,j,
\end{cases}\\
v_{jk,j-i}&=\begin{cases}
\bar\varphi_0E_{kk}\bar\sigma_{j,0}^{-1}, &i=0,\\
\Big(\bar\varphi_i-\sum_{a=0}^{i-1}\bar v_{jk,j-a}\bar\sigma_{j-a,i-a}\Big)E_{kk}\bar\sigma_{j-i,0}^{-1},& i=1,\dots,j-1,\end{cases}
  \end{align*}
and
\begin{align}
\sigma_{j,i}&:=\sum_{r=0}^{i-1}\binom{j}{r}(\partial^r\varphi_{i-r}),\\
\bar\sigma_{j,i}&:=\sum_{r=0}^{i}\binom{j}{r}(\bar\partial^r\bar\varphi_{i-r}).
\end{align}

\end{theorem}

Observe that $\sigma_{j1}=\beta$ and $\bar\sigma_{j,0}=\Exp{\phi}$,
and also that the first of the differential operators $Q_{jk}$ and $\bar
Q_{jk}$ are given by
\begin{align*}
Q_{jk}&=E_{kk}\partial^j+[\beta,E_{kk}]\partial^{j-1}
+([\varphi_2,E_{kk}]-jE_{kk}\partial\beta-[\beta,E_{kk}\beta])\partial^{j-2}+\cdots+u_{jk,0},\\
 Q_{j\bar k}&=\Exp{\phi}E_{kk}\Exp{-\phi}\bar\partial^j+(\bar\varphi_1E_{kk}-\Exp{\phi}E_{kk}
\Exp{-\phi}(\bar\varphi_1+j\bar\partial\Exp{\phi}))\Exp{-\phi}\bar\partial^{j-1}+\dots+
v_{jk,1}\bar\partial.
\end{align*}

\begin{lemma}\label{lema}
The only differential operators $Q=u_j\partial^j+\dots+u_0$ and $\bar
Q=v_j\bar\partial^j+\dots+v_0$ such that
\begin{align*}
Q(W)&=0,&
\bar Q(\bar W)&=0
\end{align*}
are
\[
Q=\bar Q=0.
\]
\end{lemma}
\begin{proof}
\begin{itemize}
\item  Let us suppose that $\sum_{i=0}^j u_i \partial^iW=0$ but $
    \sum_{i=0}^ju_i \partial^i W=\big(\sum_{i=0}^j
    (u_i+\sum_{r=1}^{j-i}u_{i+r}\sigma_{i+r,r})\Lambda^{i}+\g_-)\big)\E$.
    Thus, $u_j=u_{j-1}=\dots=u_0=0$.
\item  Assume now that $\sum_{i=1}^jv_i \bar\partial^i\bar W=0$
    and take into account that
\[
\sum_{i=1}^jv_i \bar\partial^i\bar W=\big(\sum_{i=1}^j
(\sum_{r=0}^{j-i}v_{i+r}\bar\sigma_{i+r,r})\Lambda^{-i}+\g_+\big)\bar\E.
\]
 Thus, $v_j=v_{j-1}=\dots=v_1=0$.
\end{itemize}
\end{proof}

From this lemma it follows that
\begin{pro}
The Zakharov--Shabat conditions
\begin{align*}
&\partial_{jk}Q_{il}-\partial_{il}Q_{jk}+[Q_{il},Q_{jk}]=0,&
&\partial_{j\bar k} Q_{i\bar l}- \partial_{i\bar l} Q_{j\bar k}+[ Q_{i\bar l}, Q_{j \bar k}]=0
\end{align*}
hold.
\end{pro}

\begin{proof}
Just consider the compatibility conditions of \eqref{bakerevol1}
together with Lemma \ref{lema}.
\end{proof}

The site independent relations described in Theorem \ref{pro-baker}
constitute the $N$-wave hierarchy, for the  non bared flows, and its
modification for the bared flows. These multicomponent equations
contains may integrable systems \cite{kono-oevel}, for $N=1$ we have
the KP equation in nonbared variables and the modified KP equation in
bared variables, for $N=2$, the Davey--Stewartson equation in the
$t$-variables and the Ishimori equation in the $\bar t$-variables. For
$N=3$ we find the 3-resonant wave system ($t$-variables) and a
modified version of it ($\bar t$-variables).

\begin{pro}\label{Nwave}
 The $N$-wave equations
  \[
 \partial_{1k}[\beta,E_{ll}]-\partial_{1l}[\beta,E_{kk}]+[[\beta,E_{ll}],[\beta,E_{kk}]]=0.
 \]
 and the modified $N$-wave equations
 \[
\bar\partial_{1k}(v_l)-\bar\partial_{1l}(v_k)+
v_l\bar\partial(v_k)-v_k\bar\partial(v_l)=0.
\]
with $v_k:=\bar\varphi_0 E_{kk}\bar\varphi_0^{-1}$  are satisfied.
\end{pro}
\begin{proof}
 The $N$-wave equations  appear as the compatibility of the $Q_{1k}=E_{kk}\partial_1+
 [\beta,E_{kk}]$. A ``modified" $N$-wave system \cite{kono-oevel} appears for when one considers the compatibility $\bar Q_{1k}=v_k\bar\partial $, $k=1,\dots,N$.
\end{proof}

\paragraph{Discrete versions of the modified  $N$-wave equations}

For a fixed $l=1,\dots,N$ let us introduce the following shift operator
\begin{align*}
  \bar T&:=\sum_{\substack{\bar k=\bar 1,\dots,\bar N\\\bar k\neq \bar l}}T_{(\bar k,\bar l)},
\end{align*}
and the operator
\begin{align*}
  \bar  X(A)&:=\sum_{k\neq l}T_{(\bar k,\bar l)}(A)E_{kk},&
  P_l&:=\sum_{k\neq l}E_{kk}, & A\in \g.
\end{align*}
Finally we also introduce,
\begin{align*}
\Omega_k&:=VE_{kk}V^{-1}, & V:=E_{ll}+\bar X(\Exp{\phi}),
\end{align*}
and the difference operators
\begin{align*}
  \Delta_{(\bar k,\bar l)}&:=T_{(\bar k,\bar l)}-1,&\bar\Delta&:=\sum_{k\neq l}\Delta_{(\bar k,\bar l)}=\bar T-(N-1).
\end{align*}

\begin{pro}
  The dressing operators $W$ and $\bar W$ satisfy
  \begin{align*}
    \Delta_{(\bar k,\bar l)}( W)&=\Omega_k \bar \Delta( W), & \Delta_{(\bar k,\bar l)}(\bar W)&=\Omega_k \bar \Delta (\bar W).
  \end{align*}
\end{pro}

\paragraph{Conjugate nets and quadrilateral lattices}
We show now the role of conjugate nets and quadrilateral lattices as a
part of the multicomponent Toda hierarchy.  For this aim we first prove
\begin{pro}\label{pera}
If $\epsilon_i\in M_N(\C)$, $i=1,2$ are such that $\epsilon_1
E_{kk}=\epsilon_2 E_{ll}$ then
\begin{align*}
\epsilon_1(\partial_{1k}-\beta E_{kk})W&=\epsilon_2 (T_K
-(T_K\beta E_{ll}+\I_N-E_{ll}-\pi_a))W,& K&=(l,a)\\
\epsilon_1(\partial_{1k}-\beta E_{kk})\bar W&=\epsilon_2 (T_K
-(T_K\beta E_{ll}+\I_N-E_{ll}-\pi_a)\bar W,& K&=(l,a)\\
\epsilon_1\Exp{-\phi}\partial_{1\bar k}W&=
\epsilon_2 \Exp{-T_K\phi} (\Delta_K+\pi_a) W,& K&=(\bar l,a)\\
\epsilon_1\Exp{-\phi}\partial_{1\bar k}\bar W&=
\epsilon_2 \Exp{-T_K\phi} (\Delta_K+\pi_a) \bar W,& K&=(\bar l,a).
\end{align*}
\end{pro}
A particular consequence is
\begin{multline*}
   \epsilon_1(\varphi_2E_{kk}+\partial_{1k}\varphi_1-\varphi_1E_{kk}\varphi_1)=
   \epsilon_2(T_{(k,l)}\varphi_2E_{kk}+T_{kl}\varphi_1(\I_N-E_{ll}-E_{kk})\\+E_{ll}      -(T_{kl}\varphi_1E_{kk}+\I_N-E_{ll}-E_{kk})\varphi_1).
 \end{multline*}
 If we right multiply this relation by $E_{m'm'}$ with $m'\neq k$ and we take
 \begin{enumerate}
 \item $\epsilon_1=E_{mm}$ with $m\neq k$ and $\epsilon_2=0$
 \item $\epsilon_1=0$ and $\epsilon_2=E_{mm}$ with $m\neq k,l$
 \item $\epsilon_1=0$ and $\epsilon_2=E_{ll}$
 \item $\epsilon_1=\epsilon_2=E_{kk}$
  \end{enumerate}
we find
\begin{align*}
 \partial_{1k}\beta_{mm'}-\beta_{mk}\beta_{km'}&=0,& &\text{for $m,m'\neq k$}\\
 \Delta_{(k,l)}\beta_{mm'}-(T_{(k,l)}\beta_{mk})\beta_{km'}&=0,& &\text{for $m,m'\neq k,l$}, \\
 (T_{(k,l)}\beta_{mk})\beta_{kl}+\beta_{ml}&=0,& &m\neq k,l,\\
T_{(k,l)}\beta_{lm'}-(T_{(k,l)}\beta_{lk})\beta_{km'}&=0, & &m'\neq k,l,\\(T_{(k,l)}\beta_{lk})\beta_{kl}-1&=0,&&\\
 \partial_{1k}\log\beta_{km'}-\frac{T_{(k,l)}\beta_{km'}}{\beta_{km'}}+\Delta_{(k,l)}\beta_{kk}&=0,& &m'\neq l,\\
 \partial_{1k}\log\beta_{kl}+\Delta_{(k,l)}\beta_{kk}&=0.&&
 \end{align*}

\paragraph{The dispersionfull Toda--Whitham hierarchy}

We fix  $l\in\mathbb S$ and consider the shifts $T_{(a,l)}$ for
$a\in\mathcal S$ with $a\neq l$,
 and as we can not put $a=l$  for $a'\in\mathcal S$, $a'\neq l$,
\begin{pro}\label{first TT.1}
\begin{enumerate}
\item  For $a',l$, $a'\neq l$, there exists scalar  operators
  \begin{align*}
    \B_{jl}&=T_{(l,a')}^{j}+B_{jl,j-1}T_{(l,a')}^{-j+1}+\dots+B_{jl,0},\\
\alpha_{jl}&=\partial_{1l}^j+\alpha_{jl,j-2}\partial_{1l}^{j-2}+\dots+
    \alpha_{jl,0}
  \end{align*}
  where the coefficients $B_{jl,i}$ and $\alpha_{jl,i}$ are scalar
  polynomials in the $T_{(l,a')}$-shifts or the
  $\partial_{1l}$-derivatives
   of $\beta_{ll},\varphi_{2,ll}\dots,\varphi_{j,ll}$, respectively, for
   example $B_{jl,j-1}=\beta_{ll}-T_{(l,a')}^j\beta_{ll}$ such that
  \begin{align}
    \partial_{jl}(E_{ll}W)&=\B_{jl}(E_{ll}W)=\alpha_{jl}(E_{ll}W),&  \partial_{jl}(E_{ll}\bar W)&=\B_{jl}(E_{ll}\bar W)=\alpha_{jl}(E_{ll}\bar W),\label{ttll}
  \end{align}
  \item For $a\neq l$ there exists scalar operators
  \begin{align*}
     \B_{ja}=B_{ja,j}T_{(a,l)}^j+\dots+B_{ja,1}T_{(a,l)}
  \end{align*}
  where $B_{ja,i}$ are scalar polynomials in the $T_{(l,a)}$-shifts of
  $\beta_{lk},\varphi_{2,lk},\dots,\varphi_{j,lk}$ when $a=k$ and of
  $\bar\varphi_{0,lk},\dots,\bar\varphi_{j-1,lk}$ for $a=\bar k$, for
  example
  \begin{align*}
  B_{ja,j}=\begin{cases}
    \dfrac{\beta_{lk}}{T^j_{(k,l)}(\beta_{lk})}, & a=k\in\mathbb S,\\[15pt]
  \dfrac{\bar\varphi_{0,lk}}{T^j_{(\bar k,l)}(\bar\varphi_{0,lk})}, & a=\bar k\in\bar{\mathbb S},
    \end{cases}
  \end{align*}
 such that for $a\neq l$
  \begin{align*}
    \partial_{ja}(E_{ll}W)&= \B_{ja}(E_{ll}W)&  \partial_{ja}(E_{ll}\bar W)&= \B_{ja}(E_{ll}\bar W).
  \end{align*}
\end{enumerate}
\end{pro}

\section*{Appendix B: Proofs of Propositions}
\begin{itemize}
  \item{\textbf{Proposition \ref{T-pro}}} Obviously \eqref{T-relations}
      is implied by \eqref{T-abelian}-\eqref{T-cohomological}. It is also
      easy to conclude that \eqref{T-inverse} and
      \eqref{T-cohomological} follow from \eqref{T-relations}. The non
      trivial part of the proposition is to prove that \eqref{T-relations}
      implies \eqref{T-abelian}:
\begin{align*}
 T_{(a,b)}T_{(c,d)}&= T_{(a,c)}T_{(c,b)}T_{(c,b)}T_{(b,d)}=T_{(a,c)}T_{(c,b)}T_{(b,d)}T_{(c,b)}\\
&=T_{(a,c)}T_{(c,d)}T_{(c,b)}=T_{(c,d)}T_{(a,c)}T_{(c,b)}=T_{(c,d)}T_{(a,b)}.
\end{align*}
    \item{\textbf{Proposition \ref{T-pro2}}} We only need to show that
        \eqref{con-omega} implies \eqref{con-omega1} as the reverse
        is evident. We proceed as in the proof of Proposition \ref{T-pro}
\begin{align*}
(T_{(a,b)}\omega_{(c,d)})\omega_{(a,b)}&=(T_{(a,b)}(T_{(b,d)}\omega_{(c,b)})
\omega_{(b,d)}))(T_{(a,c)}\omega_{(c,b)})\omega_{(a,c)}\\
&=(T_{(a,b)}(T_{(b,d)}\omega_{(c,b)}))(T_{(a,c)}(T_{(c,b)}
\omega_{(b,d)})\omega_{(c,b)})\omega_{(a,c)}
\\
&=(T_{(a,b)}(T_{(b,d)}\omega_{(c,b)}))(T_{(a,c)}\omega_{(c,d)})\omega_{(a,c)}
\\
&=(T_{(c,d)}(T_{(a,c)}\omega_{(c,b)}))(T_{(c,d)}\omega_{(a,c)})\omega_{(c,d)}\\
&=(T_{(c,d)}(T_{(a,c)}\omega_{(c,b)})\omega_{(a,c)})\omega_{(c,d)}
\\
&=(T_{(c,d)}\omega_{(a,b)})\omega_{(c,d)}.
\end{align*}
    \item{\textbf{Proposition \ref{lax then zs}}} We do not prove the
        differential case and refer the reader to \cite{ueno-takasaki}.
        Therefore we proceed to the remaining cases involving discrete
        times.
  \begin{enumerate}

        \item We start by proving \eqref{Omega-omega}. First, from
            the definition \eqref{omega} we deduce that
            \begin{align*}
            \partial_{ja}\omega_K\cdot\omega_K^{-1}&= \partial_{ja}(\u_K\cdot\bar\u_K^{-1})_-\cdot(\u_K\cdot\bar\u_K^{-1})_-^{-1}+
            (\u_K\cdot\bar\u_K^{-1})_-\partial_{ja}\u_K\cdot\u_K^{-1}(\u_K\cdot\bar\u_K^{-1})_-^{-1}
            \\&=\partial_{ja}(\u_K\cdot\bar\u_K^{-1})_-\cdot(\u_K\cdot\bar\u_K^{-1})_-^{-1}+
            (\u_K\cdot\bar\u_K^{-1})_-(B_{ja}-\u_K B_{ja}\u_K^{-1})(\u_K\cdot\bar\u_K^{-1})_-^{-1}\\
            &= \partial_{ja}(\u_K\cdot\bar\u_K^{-1})_-\cdot(\u_K\cdot\bar\u_K^{-1})_-^{-1}+
             (\u_K\cdot\bar\u_K^{-1})_-B_{ja}(\u_K\cdot\bar\u_K^{-1})_-^{-1}-\omega_KB_{ja}\omega_K^{-1}\\
            &\text{and similarly}
            \\&=\partial_{ja}(\u_K\cdot\bar\u_K^{-1})_+\cdot(\u_K\cdot\bar\u_K^{-1})_+^{-1}+
             (\u_K\cdot\bar\u_K^{-1})_+B_{ja}(\u_K\cdot\bar\u_K^{-1})_+^{-1}-\omega_KB_{ja}\omega_K^{-1}
            \end{align*}
            so that
             \begin{align*}\partial_{ja}\omega_K\cdot\omega_K^{-1}+\omega_KB_{ja}\omega_K^{-1}&= \partial_{ja}(\u_K\cdot\bar\u_K^{-1})_-\cdot(\u_K\cdot\bar\u_K^{-1})_-^{-1}+
             (\u_K\cdot\bar\u_K^{-1})_-B_{ja}(\u_K\cdot\bar\u_K^{-1})_-^{-1}
            \\&=\partial_{ja}(\u_K\cdot\bar\u_K^{-1})_+\cdot(\u_K\cdot\bar\u_K^{-1})_+^{-1}+
             (\u_K\cdot\bar\u_K^{-1})_+B_{ja}(\u_K\cdot\bar\u_K^{-1})_+^{-1}.
             \end{align*}

             Now, using \eqref{con-omega1} and the commuting
             character of the Lax operators we get
             \begin{align*}
             T_KB_{ja}&=(\u_K\cdot\bar\u_K^{-1})_-\r_{ja}(\u_K\cdot\bar\u_K^{-1})_-^{-1}
             -T_K(\r_{ja}-\bar\r_{ja})_-\\
             &=(\u_K\cdot\bar\u_K^{-1})_+\bar\r_{ja}(\u_K\cdot\bar\u_K^{-1})_+^{-1}+
             T_K(\r_{ja}-\bar\r_{ja})_-
             \end{align*}
             and  we deduce for $
             I:=\partial_{ja}\omega_K\cdot\omega_K^{-1}+\omega_KB_{ja}\omega_K^{-1}-T_KB_{ja}
             $ the following expressions
             \begin{align*}
             I&= \partial_{ja}(\u_K\cdot\bar\u_K^{-1})_-\cdot(\u_K\cdot\bar\u_K^{-1})_-^{-1}-
             (\u_K\cdot\bar\u_K^{-1})_-(\r_{ja}-\bar\r_{ja})_-(\u_K\cdot\bar\u_K^{-1})_-^{-1}+T_K(\r_{ja}-\bar\r_{ja})_-
            \\&=\partial_{ja}(\u_K\cdot\bar\u_K^{-1})_+\cdot(\u_K\cdot\bar\u_K^{-1})_+^{-1}+
             (\u_K\cdot\bar\u_K^{-1})_+(\r_{ja}-\bar\r_{ja})_+(\u_K\cdot\bar\u_K^{-1})_+^{-1}-T_K(\r_{ja}-\bar\r_{ja})_+.
             \end{align*}
             which hold only if $I=0$, as desired.

  \item Let us now prove \eqref{con-omega1}. From
      \eqref{omega} and \eqref{dlaxtkl} we get
  \begin{align*}
    T_K\omega_{K'}&=T_K(\u_{K'}\bar\u_{K'}^{-1})_-\cdot \omega_K\u_{K'}\omega_K^{-1}=T_K(\u_{K'}\bar\u_{K'}^{-1})_+\cdot \omega_K\bar\u_{K'}\omega_K^{-1}
  \end{align*}
  or using \eqref{omega} again
   \begin{align*}
   T_K\omega_{K'}\cdot\omega_K
   &=T_K(\u_{K'}\bar\u_{K'}^{-1})_-\cdot (\u_K\bar\u_K^{-1})_-\u_K\u_{K'}=T_K(\u_{K'}\bar\u_{K'}^{-1})_+\cdot(\u_K\bar\u_K^{-1})_+\bar\u_K
    \bar\u_{K'}.
  \end{align*}
  Then, we deduce
    \begin{align*}
   ( T_K\omega_{K'}\cdot\omega_K)_+
   &=(\u_K\u_{K'})_+,&
  (T_K\omega_{K'}\cdot\omega_K)_-  &=(\bar \u_{K'}^{-1}\bar\u_{K}^{-1})_-
  \end{align*}
  Interchanging $K\leftrightarrow K'$ and recalling the commuting
character of the Lax operators we get the desired result.
  \end{enumerate}

\item{\textbf{Proposition \ref{matrix structure}}} Let $a_{ij}$ denote
    the elements of the bi-infinite matrix $g_{k_1k_2}$, we now
    proceed to analyze the meaning of \eqref{periodic}   in different
    situations:
\begin{itemize}
\item  \textbf{Block Hankel case} Let us assume that
    $\ell_{k_1}\ell_{\bar k_2}>0$. In particular let us discuss the
    case where both integers are positive.
If we start from the element $a_{ij}$ the equation \eqref{periodic}
says that it is equal to some other element. To determine the this
element in the matrix we observe that \eqref{periodic} requires to
move in the $i$-th row $\ell_{k_1}+\ell_{\bar k_2}$ positions to
the right and in the
 diagonal passing through that position go left $\ell_{k_1}$
 positions on this diagonal, i.e. go up $\ell_{k_1}$ positions and to
 the left also $\ell_{k_1}$ positions. This gives us
  the block structure over off-diagonals as illustrated below.

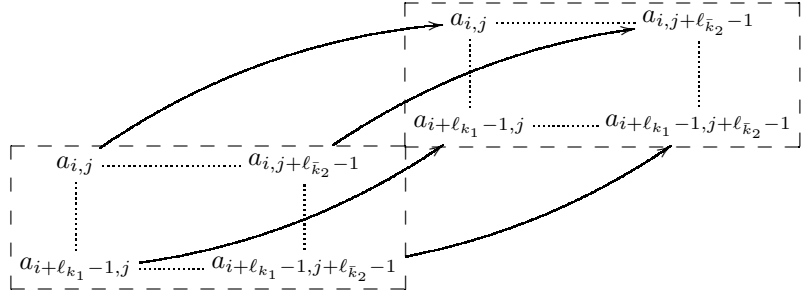
\vspace*{1cm} \hspace*{4cm} \xy \xymatrix"*"{
    a_{i,j}\ar@{.}[r]\ar@{.}[d]&a_{i,j+\ell_{\bar k_2}-1}\ar@{.}[d]\\
          a_{i+\ell_{k_1}-1,j}\ar@{.}[r]&a_{i+\ell_{k_1}-1,j+\ell_{\bar
k_2}-1} }
\POS*\frm{--}\\
\POS-(52.4,19) \xymatrix{
    a_{i,j}\ar@{.}[r]\ar@{.}[d]\ar@/^1pc/["*"]&a_{i,j+\ell_{\bar k_2}-1}\ar@{.}[d]\ar@/^1pc/["*"]\\
          a_{i+\ell_{k_1}-1,j}\ar@{.}[r]\ar@/_1pc/["*"]&a_{i+\ell_{k_1}-1,j+\ell_{\bar
k_2}-1}\ar@/_1pc/["*"] } \POS*\frm{--}
\endxy

\vspace*{1cm}

For negative integers $\ell_{k_1},\ell_{\bar k_2}<0$ we have a
 similar discussion, replacing right motions in the row with left
 motions and up motions in the diagonal with down motions in the
 diagonal,
and the same block Hankel structure appears.

\item \textbf{Block Toeplitz case} We now assume that
    $\ell_{k_1}\ell_{\bar k_2}<0$. Suppose that $\ell_{k_1}$ is
    positive. Then, when $\ell_{k_1}>|\ell_{\bar k_2}|$ if we start
    from the element $a_{ij}$ the equation \eqref{periodic} says
    that it is equal to some other element, say $a_{i'j'}$.
 To determine the row $i'$ and column $j'$, we observe that
\eqref{periodic} tell us to advance in the $i$-th row
$\ell_{k_1}-|\ell_{\bar k_2}|$ positions to the right and in the
 diagonal passing through that position go left $\ell_{k_1}$
 positions on this diagonal, i.e. go up $\ell_{k_1}$ positions and to
 the left also $\ell_{k_1}$ positions. So that we have
  \begin{align*}
   a_{ij}=a_{i-\ell_{k_1},j+\ell_{k_1}-|\ell_{\bar k_2}|-\ell_{k_1}}=a_{i-\ell_{k_1},j-|\ell_{\bar k_2}|}.
 \end{align*}
 For the case $\ell_{k_1}<|\ell_{\bar k_2}|$ we use
  $g_{j,k_1k_2}(n)=g_{j+|\ell_{\bar
  k_2}|-\ell_{k_1},k_1k_2}(n+\ell_{k_1})$ so that we move
  $|\ell_{\bar k_2}|-\ell_{k_1}$ positions to the right and  on the
  diagonal $\ell_{k_1}$ positions down, which amounts to
  $\ell_{k_1}$ rows down and $\ell_{k_1}$ columns right, i.e.
    \begin{align*}
   a_{ij}=a_{i-\ell_{k_1},j+|\ell_{\bar k_2}|-\ell_{k_1}+\ell_{k_1}}=a_{i-\ell_{k_1},j-|\ell_{\bar k_2}|}
 \end{align*}
 and we get the same result,  which immediately tell us about the
 block structure over diagonals as illustrated below.

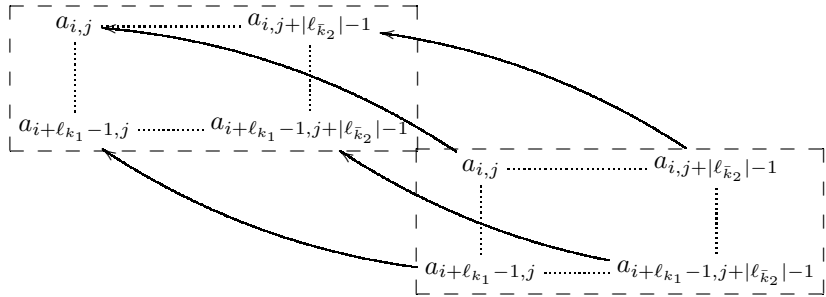
\vspace*{1cm}\hspace*{4cm}\xy \xymatrix"*"{
    a_{i,j}\ar@{.}[r]\ar@{.}[d]& a_{i,j+|\ell_{\bar k_2}|-1}\ar@{.}[d]\\
      a_{i+\ell_{k_1}-1,j}   \ar@{.}[r]&a_{i+\ell_{k_1}-1,j+|\ell_{\bar
k_2}|-1} }
\POS*\frm{--}\\
\POS-(-54,19) \xymatrix{
    a_{i,j}\ar@{.}[r]\ar@{.}[d]\ar@/_1pc/["*"]& a_{i,j+|\ell_{\bar k_2}|-1}\ar@{.}[d]\ar@/_1pc/["*"]\\
      a_{i+\ell_{k_1}-1,j}
\ar@{.}[r]\ar@/^1pc/["*"]&a_{i+\ell_{k_1}-1,j+|\ell_{\bar
k_2}|-1}\ar@/^1pc/["*"] } \POS*\frm{--}
\endxy
\vspace*{1cm}

A similar discussion goes on for the case of negative $\ell_{k_1}$
and positive $\ell_{\bar k_2}$.

\item The case $\ell_{k_1}=0$ with $\ell_{\bar k_2}\neq 0$
    gives $g_{j,k_1k_2}(n)=g_{j+\ell_{\bar k_2},k_1k_2}(n)$, which
    implies diagonal band structure, whether for $\ell_{\bar
    k_2}=0$ with $\ell_{ k_1}\neq 0$ gives
    $g_{j,k_1k_2}(n)=g_{j+\ell_{k_1},k_1k_2}(n+\ell_{k_1})$,
    which describes a $\ell_{k_1}\times\ell_{k_1}$ block structure.
    Notice that these two cases can only exist for two or more
    components.
\end{itemize}

\item{\textbf{Proposition \ref{pro:orlov}}} Let us compute
  \begin{align}
\label{m01}    M&=W nW^{-1}=S\E n\E^{-1} S^{-1},\\
\label{barm01}    \bar M&=\bar W n\bar W^{-1}=\bar S\bar\E n\bar\E^{-1} \bar S^{-1},
  \end{align}
  for this aim we must take into account that
\begin{gather}\label{AdE}
\begin{aligned}
 \mu:=\E n\E^{-1}&=n+\nu,& \nu&=\sum_{k=1}^NE_{kk}(s_k+\sum_{j=1}^\infty j t_{jk}\Lambda^j),\\
 \bar\mu:=\bar\E n\bar\E^{-1}&=n+\bar\nu,&
  \bar\nu&=-\sum_{k=1}^NE_{kk}( s_{\bar k}+\sum_{j=1}^\infty j
  t_{j\bar k}\Lambda^{-j}).
  \end{aligned}
\end{gather}

Therefore, from \eqref{m01} and \eqref{barm01} we deduce that
\begin{align*}
  M&= S\mu S^{-1}=SnS^{-1}
  +\sum_{k=1}^N C_{kk}(s_k+\sum_{j=1}^\infty j t_{jk}L^j),\\
  \bar M &=\bar S\bar\mu\bar S^{-1}=\bar Sn\bar S^{-1}
  -\sum_{k=1}^N \bar C_{kk}( s_{\bar k}+\sum_{j=1}^\infty j  t_{j\bar k}\bar L^{-j}).
\end{align*}
Finally, 
\begin{align*}
  \m&:=SnS^{-1}=(1+\beta(n)\Lambda^{-1}+\varphi_2(n)\Lambda^{-2}+\cdots)n (1+\beta(n)\Lambda^{-1}+\varphi_2(n)\Lambda^{-2}+\cdots)^{-1}\\&=
  n-\beta(n)\Lambda^{-1}+\cdots,\\
  \bar \m&:=\bar S n\bar S^{-1}=(\Exp{\phi(n)}+\bar\varphi_1(n)\Lambda+\cdots)n
  (\Exp{\phi(n)}+{\bar\psi}_1(n)\Lambda+\cdots)^{-1}\\&=
  n+\bar\varphi_1(n)\Exp{-\phi(n+1)}\Lambda+\cdots.
\end{align*}
\item{\textbf{Proposition \ref{undressing Ckl}}} Let us take $W$ of
    Theorem \ref{undressing lax orlov}  and consider
$    \Theta_{kl}:=W^{-1}C_{kl}W$ which satisfy
$    [\Theta_{kl},\Lambda]=[\Theta_{kl},n]=0$ and hence
  $\Theta_{kl}$    do not depends on $\Lambda$ nor on $n$. Now,
  \begin{align*}
    E_{k'k'}\Theta_{kl}&=\delta_{k'k}\Theta_{k'l},\quad \Theta_{kl}E_{k'k'}=\delta_{lk'}\Theta_{kl}\quad\Rightarrow\quad\Theta_{kl}=\vartheta_{kl}E_{kl},
    \quad\vartheta_{kl}\in\C,\\
       E_{k'k'}\bar\Theta_{kl}&=\delta_{k'k}\bar\Theta_{k'l},\quad \bar  \Theta_{kl}E_{k'k'}=\delta_{lk'}\bar\Theta_{kl}\quad\Rightarrow\quad\bar\Theta_{kl}=\bar\vartheta_{kl}E_{kl},
   \quad\bar\vartheta_{kl}\in\C.
  \end{align*}
  Thus,
  \begin{multline*}
    C_{kl}=W\Theta_{kl}W^{-1}=SW_0\vartheta_{kl}E_{kl}W_0^{-1}S^{-1}=
    \vartheta_{kl}L^{s_k-s_l}\Exp{\sum_{j\geq
    1}(t_{jk}-t_{jl})L^j}(E_{kl}+\g_-)\\\Rightarrow\vartheta_{kl}=1\Rightarrow
    C_{kl}=WE_{kl}W^{-1}.
  \end{multline*}
\end{itemize}

\section*{Acknowledgements}
The authors wish to thank the Spanish Ministerio de Ciencia e
Innovaci\'{o}n, research projects FIS2005-00319 and FIS2008-00200,
and  acknowledge the support received from the European Science
Foundation (ESF) and the activity entitled \emph{Methods of Integrable
Systems, Geometry, Applied Mathematics} (MISGAM). This paper was
finished during the research visits of one of the authors (MM) to the
\emph{ Universit\'{e} Catholique de Louvain } and to the \emph{Scuola
Internazionale Superiore di Studi Avanzati}/International School for
Advanced Studies
 (SISSA) in Trieste, MM wish to thanks Prof. van Moerbeke and Prof. Dubrovin for their warm hospitality, acknowledge
 economical support from MISGAM and  SISSA  and
 reckons different conversations with P. van Moerbeke,  T. Grava, G. Carlet and M. Caffasso.

\end{document}